\documentclass[prd,preprintnumbers,amsmath,amssymb,nofootinbib]{revtex4}
\usepackage{amsmath}
\usepackage{footnote}
\usepackage{graphicx}
\usepackage{multirow}
\usepackage{diagbox}
\usepackage{xcolor} 
\usepackage{cancel}
\usepackage[figuresleft]{rotating}
\usepackage{booktabs}
\usepackage{appendix}
\usepackage{multirow}
\usepackage{slashed}
\usepackage{xcolor}
\usepackage{graphicx}
\usepackage{epstopdf}
\usepackage{ulem}
\usepackage[colorlinks,
            citecolor=blue,
            anchorcolor=red,
            menucolor=red,
            linkcolor=red,
            filecolor=red,
            runcolor=red,
            urlcolor=blue,
            frenchlinks=red]{hyperref}
\usepackage{cleveref}
\begin{document}
\title{Magnetic Moments and Radiative Transitions of the $T_{cc}(3875)^+$ and its partner states}
\author{Jing Wu$^1$\footnote{Contact authors: wujing18@sdjzu.edu.cn; yrliu@sdu.edu.cn; }}
\author{Yi-Kun Wang$^1$}
\author{Kai-Bao Chen$^1$}
\author{Yan-Rui Liu$^2$\footnotemark[\value{footnote}]}
\author{Jian-Bo Cheng$^3$}
\author{Wei-Hua Yang$^4$}
\affiliation{
    $^1$School of Science, Shandong Jianzhu University, Jinan 250101, China\\
	$^2$School of Physics, Shandong University, Jinan, Shandong 250100, China\\
    $^3$College of Science, China University of Petroleum, Qingdao, Shandong 266580, China\\
	$^4$Department of Nuclear Physics, Yantai University, Yantai, Shandong 264005, China
}
   
\date{\today}

\begin{abstract}
We systematically investigate the magnetic moments (MMs) and radiative decay widths of S-wave doubly heavy tetraquark states based on chromomagnetic interaction (CMI). They provide a complementary probe to distinguish compact from molecule structures in addition to the mass spectrum and strong decay properties.
Using the CMI eigenvectors, we compute the MMs and M1 transition rates for the tetraquarks in the compact configuration. We predict the MM of the observed $T_{cc}(3875)^+$ with $I(J^P)=0(1^+)$ to be 0.45 $\mu_N$ when treating it as a compact tetraquark state, whereas the MM of the $0(1^+)$ $DD^*$ molecule is about $-0.07\,\mu_N$. Five radiative transition channels related to the $T_{cc}(3875)^+$ are identified with widths ranging from $6.07$ keV to $306.37$ keV.
Our results show that the MMs of the $J^P=1^+$ $bc\bar{q}\bar{q}^\prime$ ($q/q^\prime=u,d,s$) and $J^P=1^+$ $QQ\bar{n}\bar{s}$ ($Q=b,c; n=u,d$) states are influenced by the diquark-spin mixing.
Through the analyses of radiative transitions between different tetraquark states, we find that such processes in the $QQ\bar{n}\bar{n}^\prime$, $QQ\bar{s}\bar{s}$, and $bc\bar{n}\bar{s}$ cases may serve to reveal the tetraquark structures of the initial or final states.
We also define the magnetic coupling matrices characterizing the MMs of the tetraquark system with $J^P=1^+$, with which the range of MM can be constrained. The present study provides a valuable reference point for the search of exotic states in future particle physics experiments.
\end{abstract}
\maketitle

\section{Introduction}\label{sec1}

The discovery of the doubly charmed tetraquark state $T_{cc}(3875)^+$ by the LHCb Collaboration in the $D^{*+}D^0$ invariant mass spectrum \cite{LHCb:2021vvq,LHCb:2021auc} provides the first unambiguous evidence for a narrow exotic hadron with the minimal quark content $cc\bar{u}\bar{d}$. Its quantum number, mass, and width are $I(J^P)=0(1^+)$, $3875.01\pm0.26\pm0.17$ MeV, and $410\pm165^{+18}_{-38}$ keV, respectively. This state serves as an ideal laboratory for studying the non-perturbative quantum chromodynamics (QCD) dynamics. Since the $T_{cc}(3875)^+$ state just lies $273\pm61\pm5^{+11}_{-14}$ keV below the $D^{*+}D^0$ threshold, its properties may be naturally understood in the molecule picture. However, the precise nature of the $T_{cc}(3875)^+$, a molecule, a compact tetraquark, or an admixture thereof, still remains an open question.

In the literature, after the observation of the $T_{cc}(3875)^+$, the molecular and compact tetraquark configurations have been widely adopted to understand its internal structure. The former configuration describes the $T_{cc}(3875)^+$ as a loosely bound state of $D$ and $D^*$ mesons, formed via the exchange of light mesons such as $\pi$, $\rho$, and $\omega$.
Studies based on effective field theories \cite{Zhang:2025rlg,Xu:2025xrl,Meng:2023bmz,Abolnikov:2024key,Collins:2024sfi,Kinugawa:2023fbf,Yan:2021wdl,Fleming:2021wmk}, one-boson-exchange models \cite{Lu:2025zae,Sun:2024wxz,Chen:2021vhg,Wang:2024vjc,Deng:2021gnb}, and lattice QCD simulations \cite{Nagatsuka:2025szy,Whyte:2024ihh,Stump:2025owq} can successfully reproduce its near-threshold characteristics and the tiny binding energy. 
Conversely, the latter configuration treats the $T_{cc}(3875)^+$ as a tightly bound state composed of a diquark and an antidiquark, and yields mass predictions consistent with experimental observations \cite{Agaev:2021vur,Prelovsek:2025vbr,Guo:2021yws,Weng:2021hje}. In effect, numerous theoretical studies had already explored the possible existence of bound states in the $QQ \bar{q}\bar{q}^\prime$ system (with $Q=c,b$ and $q/q^\prime=u,d,s$) long before the experimental discovery of the $T_{cc}(3875)^+$ state. One may find a comprehensive review in the literature \cite{Liu:2019zoy}. For convenience, in the following discussions, we usually use $n$ to denote the $u$ or $d$ quark.

Complementary to the mass spectrum, the strong decay properties \cite{Sun:2024wxz,Jia:2022qwr,Fleming:2021wmk,Yan:2021wdl,Ling:2021bir} and the $DD\pi$ three-body effects \cite{Lin:2022wmj,Feijoo:2021ppq,Guo:2023xyf} may also be indicative of the internal structure of $T_{cc}(3875)^+$. 
The final-state interactions, triangle singularities, and the scattering lengths  \cite{Jia:2022qwr,Dai:2023mxm,Braaten:2022elw,Achasov:2022onn,Du:2021zzh,Meng:2023bmz} provide additionally important information for understanding the nature of the $T_{cc}(3875)^+$. These mentioned studies consistently point to an extremely narrow width (tens of keV), supporting its molecule interpretation.

Beyond the spectrum and the strong decay properties, the electromagnetic properties may provide further insights into the internal structure of the $T_{cc}(3875)^+$. Theoretical studies \cite{ Zhang:2025wmr, Lei:2023ttd, Azizi:2021aib, Meng:2021jnw,Deng:2021gnb,Zhang:2021yul} indicate that the observables such as magnetic moments (MMs) and radiative decay widths are highly sensitive to the internal charge and spin distributions and they may serve as ideal probes to understand the $T_{cc}(3875)^+$. 
However, the MM values predicted by different models exhibit notable differences, from $-0.09$ $\mu_N$ to $0.43$ $\mu_N$ for the molecule interpretation and $-0.28$ $\mu_N$ to $0.76$ $\mu_N$ for the compact tetraquark interpretation.
Even within the framework of the same structural model, different theoretical approaches may yield discrepant predictions for the MM of a doubly charmed tetraquark state. To facilitate future experimental tests of theoretical frameworks, it is imperative to carry out systematic investigations, from multiple perspectives, into the electromagnetic properties of doubly heavy tetraquark states.

Motivated by the above considerations, we calculate the magnetic moments of doubly heavy tetraquark states in this paper.
However, the internal structure of the S-wave $J=0$ states cannot be probed via their vanishing MMs.
As a complementary analysis, we also investigate the radiative transitions between different doubly heavy tatraquark states. Such radiative decays can not only reveal the structural information of the $J=0$ states, but also uncover the intrinsic correlations among different tetraquarks.
The MMs and radiative decay widths of exotic states can be used to distinguish between molecular and compact tetraquark models, particularly for those states lying near meson-meson thresholds.

This paper is arranged as follows. In Sec. \ref{sec2}, we present the mass splitting model used to study the magnetic moments and radiative decays. The wave functions, chromomagnetic interaction (CMI) matrices, and expressions for magnetic moments and radiative decay widths are provided. The numerical results are given and discussed in Sec. \ref{sec3}. Finally, we present the summary in Sec. \ref{sec4}.

\section{The model}\label{sec2}

When the nonrelativistic quark model is applied to S-wave doubly heavy tetraquarks, both the tensor and spin-orbit terms of the potential are negligible. For our purposes, we need the spin-color wave function of a tetraquark. It is convenient to employ the CMI model where the spatial wave function does not explicitly appear. The Hamiltonian is reduced to
\begin{eqnarray}
    &\hat{H}=\sum_i m_i+H_{CMI},&\\
    &H_{CMI}=-\sum_{i<j}C_{ij}\lambda_i\cdot\lambda_j\sigma_i\cdot\sigma_j\label{equ:CMI},&
\end{eqnarray}
where $m_i$ denotes the effective mass of the $i$-th quark, $H_{CMI}$ is the chromomagnetic interaction.
$\lambda_i$ and $\sigma_i$ are the Gell-Mann matrix and the Pauli matrix for the $i$-th quark, respectively.
Since the S-wave spatial contribution is already absorbed into the effective coupling constant $C_{ij}$ which represents the interaction strength between quarks $i$ and $j$, we focus on the remaining color–spin components of the wave function subject to the Pauli principle.
We use $|\psi^{YIJ}_{Q_1Q_2,i}\rangle=[(Q_1Q_2)_{color}^{s_{12}}(\bar{q_3}\bar{q_4})_{color}^{s_{34}}]^J_{1_c}$ to describe the flavor, spin, and color structure of the doubly heavy tetraquark state characterized by the hypercharge ($Y$), isospin ($I$), and total angular momentum ($J$). The basis wave functions are listed in \Cref{tab:tetraquark-wave-function}.
For a comprehensive account of the construction of these wave functions, we refer the readers to our earlier studies \cite{Wu:2016gas,Chen:2016ont,Li:2023wug,Wu:2016vtq}.
\begin{table}
	\caption{Basis wave functions of the S-wave doubly heavy tetraquark states marked as $\psi^{(YIJ)}_{QQ^\prime}$ ($Q/Q^\prime=c$ or $b$). In our notation, the superscripts of diquark, anti-diquark, and tetraquark denote explicitly their spins, while the subscripts denote their color representations.}\label{tab:tetraquark-wave-function}
	\begin{tabular}{c|c|c|cccccc}\hline\hline
		System&$J=2$&$J=1$&$J=0$\\\hline
		\multirow{4}{*}{\begin{tabular}{c}$cc\bar{n}\bar{n}^\prime$\\$bb\bar{n}\bar{n}^\prime$\end{tabular}}
		&$\psi^{(-\frac2312)}_{QQ}=[(QQ)^1_{\bar{3}}(\bar{n}\bar{n}^\prime)^1_3]^2_{1_c}$&$\psi^{(-\frac2311)}_{QQ}=[(QQ)^1_{\bar{3}}(\bar{n}\bar{n}^\prime)^1_3]^1_{1_c}$&$\psi^{(-\frac2310)}_{QQ,1}=[(QQ)^1_{\bar{3}}(\bar{n}\bar{n}^\prime)^1_3]^0_{1_c}$\\
		&&&$\psi^{(-\frac2310)}_{QQ,2}=[(QQ)^0_6(\bar{n}\bar{n}^\prime)^0_{\bar{6}}]^0_{1_c}$\\
		\cline{2-4}
		&&$\psi^{(-\frac2301)}_{QQ,1}=[(QQ)^1_{\bar{3}}(\bar{n}\bar{n}^\prime)^0_3]^1_{1_c}$&\\
		&&$\psi^{(-\frac2301)}_{QQ,2}=[(QQ)^0_6(\bar{n}\bar{n}^\prime)^1_{\bar{6}}]^1_{1_c}$&\\
		\hline
		\multirow{2}{*}{\begin{tabular}{c}$cc\bar{s}\bar{s}$\\$bb\bar{s}\bar{s}$\end{tabular}}
		&$\psi^{(\frac4302)}_{QQ}=[(QQ)^1_{\bar{3}}(\bar{s}\bar{s})^1_3]^2_{1_c}$&$\psi^{(\frac4301)}_{QQ}=[(QQ)^1_{\bar{3}}(\bar{s}\bar{s})^1_3]^1_{1_c}$&$\psi^{(\frac4300)}_{QQ,1}=[(QQ)^1_{\bar{3}}(\bar{s}\bar{s})^1_3]^0_{1_c}$\\
		&&&$\psi^{(\frac4300)}_{QQ,2}=[(QQ)^0_6(\bar{s}\bar{s})^0_{\bar{6}}]^0_{1_c}$\\
		\hline
		\multirow{3}{*}{\begin{tabular}{c}$cc\bar{n}\bar{s}$\\$bb\bar{n}\bar{s}$\end{tabular}}
			&$\psi^{(\frac13\frac122)}_{QQ}=[(QQ)^1_{\bar{3}}(\bar{n}\bar{s})^1_3]^2_{1_c}$&$\psi^{(\frac13\frac121)}_{QQ,1}=[(QQ)^1_{\bar{3}}(\bar{n}\bar{s})^1_3]^1_{1_c}$&$\psi^{(\frac13\frac120)}_{QQ,1}=[(QQ)^1_{\bar{3}}(\bar{n}\bar{s})^1_3]^0_{1_c}$\\
			&&$\psi^{(\frac13\frac121)}_{QQ,2}=[(QQ)^1_{\bar{3}}(\bar{n}\bar{s})^0_3]^1_{1_c}$&$\psi^{(\frac13\frac120)}_{QQ,2}=[(QQ)^0_{6}(\bar{n}\bar{s})^0_{\bar{6}}]^0_{1_c}$\\
			&&$\psi^{(\frac13\frac121)}_{QQ,3}=[(QQ)^0_6(\bar{n}\bar{s})^1_{\bar{6}}]^1_{1_c}$&\\
			\hline
			\multirow{6}{*}{$bc\bar{n}\bar{n}^\prime$}
			&$\psi^{(-\frac2312)}_{bc}=[(bc)^1_{\bar{3}}(\bar{n}\bar{n}^\prime)^1_3]^2_{1_c}$&$\psi^{(-\frac2311)}_{bc,1}=[(bc)^1_{\bar{3}}(\bar{n}\bar{n}^\prime)^1_3]^1_{1_c}$&$\psi^{(-\frac2310)}_{bc,1}=[(bc)^1_{\bar{3}}(\bar{n}\bar{n}^\prime)^1_3]^0_{1_c}$\\
			&&$\psi^{(-\frac2311)}_{bc,2}=[(bc)^1_6(\bar{n}\bar{n}^\prime)^0_{\bar{6}}]^1_{1_c}$&$\psi^{(-\frac2310)}_{bc,2}=[(bc)^0_6(\bar{n}\bar{n}^\prime)^0_{\bar{6}}]^0_{1_c}$\\
			&&$\psi^{(-\frac2311)}_{bc,3}=[(bc)^0_{\bar{3}}(\bar{n}\bar{n}^\prime)^1_3]^1_{1_c}$&\\
			\cline{2-4}
			&$\psi^{(-\frac2302)}_{bc}=[(bc)^1_6(\bar{n}\bar{n}^\prime)^1_{\bar{6}}]^2_{1_c}$&$\psi^{(-\frac2301)}_{bc,1}=[(bc)^1_6(\bar{n}\bar{n}^\prime)^1_{\bar{6}}]^1_{1_c}$&$\psi^{(-\frac2300)}_{bc,1}=[(bc)^1_6(\bar{n}\bar{n}^\prime)^1_{\bar{6}}]^0_{1_c}$\\
			&&$\psi^{(-\frac2301)}_{bc,2}=[(bc)^1_{\bar{3}}(\bar{n}\bar{n}^\prime)^0_3]^1_{1_c}$&$\psi^{(-\frac2300)}_{bc,2}=[(bc)^0_{\bar{3}}(\bar{n}\bar{n}^\prime)^0_3]^0_{1_c}$\\
			&&$\psi^{(-\frac2301)}_{bc,3}=[(bc)^0_6(\bar{n}\bar{n}^\prime)^1_{\bar{6}}]^1_{1_c}$&\\
			\hline
			\multirow{3}{*}{$bc\bar{s}\bar{s}$}
			&$\psi^{(\frac4302)}_{bc}=[(bc)^1_{\bar{3}}(\bar{s}\bar{s})^1_3]^2_{1_c}$&$\psi^{(\frac4301)}_{bc,1}=[(bc)^1_{\bar{3}}(\bar{s}\bar{s})^1_3]^1_{1_c}$&$\psi^{(\frac4300)}_{bc,1}=[(bc)^1_{\bar{3}}(\bar{s}\bar{s})^1_3]^0_{1_c}$\\
			&&$\psi^{(\frac4301)}_{bc,2}=[(bc)^1_6(\bar{s}\bar{s})^0_{\bar{6}}]^1_{1_c}$&$\psi^{(\frac4300)}_{bc,2}=[(bc)^0_6(\bar{s}\bar{s})^0_{\bar{3}}]^0_{1_c}$\\
			&&$\psi^{(\frac4301)}_{bc,3}=[(bc)^0_{\bar{3}}(\bar{s}\bar{s})^1_3]^1_{1_c}$&\\
			\hline
			\multirow{6}{*}{$bc\bar{n}\bar{s}$}
			&$\psi^{(\frac13\frac122)}_{bc,1}=[(bc)^1_6(\bar{n}\bar{s})^1_{\bar{6}}]^2_{1_c}$&$\psi^{(\frac13\frac121)}_{bc,1}=[(bc)^1_6(\bar{n}\bar{s})^1_{\bar{6}}]^1_{1_c}$&$\psi^{(\frac13\frac120)}_{bc,1}=[(bc)^1_6(\bar{n}\bar{s})^1_{\bar{6}}]^0_{1_c}$\\
			&$\psi^{(\frac13\frac122)}_{bc,2}=[(bc)^1_{\bar{3}}(\bar{n}\bar{s})^1_3]^2_{1_c}$&$\psi^{(\frac13\frac121)}_{bc,2}=[(bc)^1_{\bar{3}}(\bar{n}\bar{s})^1_3]^1_{1_c}$&$\psi^{(\frac13\frac120)}_{bc,2}=[(bc)^1_{\bar{3}}(\bar{n}\bar{s})^1_3]^0_{1_c}$\\
			&&$\psi^{(\frac13\frac121)}_{bc,3}=[(bc)^1_6(\bar{n}\bar{s})^0_{\bar{6}}]^1_{1_c}$&$\psi^{(\frac13\frac120)}_{bc,3}=[(bc)^0_6(\bar{n}\bar{s})^0_{\bar{6}}]^0_{1_c}$\\
			&&$\psi^{(\frac13\frac121)}_{bc,4}=[(bc)^1_{\bar{3}}(\bar{n}\bar{s})^0_3]^1_{1_c}$&$\psi^{(\frac13\frac120)}_{bc,4}=[(bc)^0_{\bar{3}}(\bar{n}\bar{s})^0_3]^0_{1_c}$\\
			&&$\psi^{(\frac13\frac121)}_{bc,5}=[(bc)^0_6(\bar{n}\bar{s})^1_{\bar{6}}]^1_{1_c}$&\\
			&&$\psi^{(\frac13\frac121)}_{bc,6}=[(bc)^0_{\bar{3}}(\bar{n}\bar{s})^1_3]^1_{1_c}$&\\
			\hline\hline
	\end{tabular}
\end{table}

According to \Cref{tab:tetraquark-wave-function}, one can calculate the CMI matrix $\langle H_{CMI}\rangle$ for the S-wave doubly heavy tetraquark states. Each matrix element is given by
\begin{eqnarray}
	\langle\psi^{YIJ}_{Q_1Q_2,i}|H_{CMI}|\psi^{YIJ}_{Q_1Q_2,j}\rangle = -\sum_{k<l} C_{kl} \langle\psi^{YIJ}_{Q_1Q_2,i}| \lambda_k \cdot \lambda_l\, \sigma_k \cdot \sigma_l |\psi^{YIJ}_{Q_1Q_2,j}\rangle,
\end{eqnarray}
where the sum runs over all pairs of quark components in the system. For convenience, we collect all the CMI matrices for the $QQ^\prime  \bar{q}\bar{q}^\prime$ tetraquark states in the third column of \Cref{CMI matrices}. A detailed discussion can be found in Ref.~\cite{Li:2023wug}.
\setlength{\tabcolsep}{1.5mm}\begin{table}[htbp]
	\caption{Expressions of CMI matrices for the $QQ^\prime  \bar{q}\bar{q}^\prime$ tetraquark states where $Q/Q^\prime=c$ or $b$, $q/q^\prime=n,s$.}\scriptsize\label{CMI matrices}
	\begin{tabular}{cccc}\hline\hline
		System & $J^{P}$ &$\langle H_{CMI} \rangle$& Wave function base \\\hline
		$[cc\bar{n}\bar{n}^\prime] ([bb\bar{n}\bar{n}^\prime], cc\bar{s}\bar{s}, bb\bar{s}\bar{s})$ &$2^+$&$\frac{4}{3}(2\tau+\alpha)$&$(\psi_{cc}^{(-\frac2312)})^T$\\
		&$1^+$&$\frac{4}{3}(2\tau-\alpha)$&$(\psi_{cc}^{(-\frac2311)})^T$\\
		&$0^+$&$\begin{pmatrix}
		\frac{8}{3}(\tau-\alpha) &2\sqrt{6}\alpha\\
		&4\tau\\
		\end{pmatrix}$&$(\psi_{cc,1}^{(-\frac2310)},\psi_{cc,2}^{(-\frac2310)})^T$\\
		$[cc\bar{n}\bar{n}^\prime] ([bb\bar{n}\bar{n}^\prime])$&$1^+$&$\begin{pmatrix}
		-\frac{8}{3}\eta &2\sqrt{2}\alpha\\
		&\frac{4}{3}\gamma\\
		\end{pmatrix}$&$(\psi_{cc,1}^{(-\frac2301)},\psi_{cc,2}^{(-\frac2301)})^T$\\
		$[bc\bar{n}\bar{n}^\prime] (bc\bar{s}\bar{s})$            &$2^+$&$\frac{4}{3}(2\tau+\alpha)$&$(\psi_{bc}^{(-\frac2312)})^T$\\
		&$1^+$&$\begin{pmatrix}
		\frac{4}{3}(2\tau-\alpha) &-4\mu &-\frac{4\sqrt{2}}{3}\mu \\
		&-\frac{4}{3}\eta&-2\sqrt{2}\tau\\
		&                &-\frac{8}{3}\gamma\\
		\end{pmatrix}$&$(\psi_{bc,1}^{(-\frac2311)},\psi_{bc,2}^{(-\frac2311)},\psi_{bc,3}^{(-\frac2311)})^T$\\
		&$0^+$&$\begin{pmatrix}
		\frac{8}{3}(\tau-\alpha) &2\sqrt{6}\alpha\\
		&4\tau\\
		\end{pmatrix}$&$(\psi_{bc,1}^{(-\frac2310)},\psi_{bc,2}^{(-\frac2311)})^T$\\
		$[bc\bar{n}\bar{n}^\prime]$             &$2^+$&$\frac{2}{3}(-2\tau+5\alpha)$&$(\psi_{bc}^{(-\frac2302)})^T$\\
		&$1^+$&$\begin{pmatrix}
		-\frac{2}{3}(\tau+5\alpha) &-4\mu &-\frac{10\sqrt{2}}{3}\mu \\
		&\frac{8}{3}\eta&-2\sqrt{2}\alpha\\
		&                &\frac{4}{3}\gamma\\
		\end{pmatrix}$&$(\psi_{bc,1}^{(-\frac2301)},\psi_{bc,2}^{(-\frac2301)},\psi_{bc,3}^{(-\frac2301)})^T$\\
		&$0^+$&$\begin{pmatrix}
		-\frac{4}{3}(\tau+5\alpha) &2\sqrt{6}\alpha\\
		&-8\tau\\
		\end{pmatrix}$&$(\psi_{bc,1}^{(-\frac2300)},\psi_{bc,2}^{(-\frac2300)})^T$\\
		$cc\bar{n}\bar{s} (bb\bar{n}\bar{s})$                   &$2^+$&$\frac{4}{3}(2\tau+\alpha)$&$(\psi^{(\frac13\frac122)}_{cc})^T$\\
		&$1^+$&$\begin{pmatrix}
		\frac{4}{3}(2\tau-\alpha) &\frac{4\sqrt{2}}{3}\beta &4\beta \\
		&\frac{8}{3}\eta&-2\sqrt{2}\alpha\\
		&                &\frac{4}{3}\gamma\\
		\end{pmatrix}$&$(\psi^{(\frac13\frac121)}_{cc,1},\psi^{(\frac13\frac121)}_{cc,2},\psi^{(\frac13\frac121)}_{cc,3})^T$\\
		&$0^+$&$\begin{pmatrix}
		\frac{8}{3}(\tau-\alpha) &2\sqrt{6}\alpha\\
		&4\tau\\
		\end{pmatrix}$&$(\psi^{(\frac13\frac120)}_{cc,1},\psi^{(\frac13\frac120)}_{cc,2})^T$\\
		$bc\bar{n}\bar{s}$                      &$2^+$&$\begin{pmatrix}
		\frac{2}{3}(-2\tau+5\alpha) &2\sqrt{2}\nu\\
		&\frac{4}{3}(2\tau+\alpha)\\
		\end{pmatrix}$&$(\psi^{(\frac13\frac122)}_{bc,1},\psi^{(\frac13\frac122)}_{bc,2})^T$\\
		&$1^+$&$\begin{pmatrix}
		-\frac{2}{3}(2\tau+5\alpha) &2\sqrt{2}\nu &\frac{10\sqrt{2}}{3}\beta &-4\mu &-\frac{10\sqrt{2}}{\mu} &4\beta \\
		&\frac{4}{3}(2\tau-\alpha)&-4\mu &\frac{4\sqrt{2}}{3}\beta &4\beta &-\frac{4\sqrt{2}}{3}\mu \\
		&                &-\frac{4}{3}\eta &0  &\frac{10}{3}\nu &-2\sqrt{2}\alpha\\
		&                 &                &\frac{8}{3}\eta &-2\sqrt{2}\alpha &\frac{4}{3}\nu\\
		&                 &                &                 &\frac{4}{3}\gamma&0                     \\
		&                  &               &                 &           &-\frac{8}{3}\gamma\\
		\end{pmatrix}$&$\begin{matrix}(\psi^{(\frac13\frac121)}_{bc,1},\psi^{(\frac13\frac121)}_{bc,2},\psi^{(\frac13\frac121)}_{bc,3},\\
		\psi^{(\frac13\frac121)}_{bc,4},\psi^{(\frac13\frac121)}_{bc,5},\psi^{(\frac13\frac121)}_{bc,6})^T\end{matrix}$\\
		&$0^+$&$\begin{pmatrix}
		-\frac{4}{3}(\tau+5\alpha) &4\sqrt{2}\nu &-\frac{10}{\sqrt{3}}\nu &2\sqrt{6}\alpha\\
		&\frac{8}{3}(\tau-\alpha)&2\sqrt{6}\alpha &-\frac{4}{\sqrt{3}}\nu\\
		&2\sqrt{6}\alpha &4\tau &0\\
		&&&-8\tau\\
		\end{pmatrix}$&$\begin{matrix}(\psi^{(\frac13\frac120)}_{bc,1},\psi^{(\frac13\frac120)}_{bc,2},\\
		\psi^{(\frac13\frac120)}_{bc,3},\psi^{(\frac13\frac120)}_{bc,4})^T\end{matrix}$\\\hline\hline
		
	\end{tabular}
\end{table}
Those variables appearing in the table are defined as follows: $\tau=C_{12}+C_{34}$, $\theta=C_{12}-C_{34}$, $\alpha=C_{13}+C_{14}+C_{23}+C_{24}$, $\beta=C_{13}-C_{14}+C_{23}-C_{24}$, $\mu=C_{13}+C_{14}-C_{23}-C_{24}$, $\nu=C_{13}-C_{14}-C_{23}+C_{24}$, $\gamma=3C_{12}-C_{34}$, and $\eta=C_{12}-3C_{34}$. 
Diagonalizing the CMI matrices yields the eigenvalues and eigenvectors, which correspond to the mass shifts to some scale and the CMI eigenstates, respectively.

We adopt the CMI eigenstates to study the magnetic moments and radiative decay properties of the $QQ^\prime  \bar{q}\bar{q}^\prime$ tetraquarks. The flavor-spin-color wave function of a tetraquark is expressed as a linear combination of the basis functions $|\psi^{YIJ}_{QQ^\prime,i}\rangle$ listed in \Cref{tab:tetraquark-wave-function},
\begin{eqnarray}
    |\Psi\rangle=\sum\alpha_i|\psi^{YIJ}_{QQ^\prime,i}\rangle\label{func:tetraquark},
\end{eqnarray}
where $\alpha_i$ denotes the eigenvector element of the $\langle H_{CMI}\rangle$ matrix, representing the weight coefficient of each spin-color configuration in the compact tetraquark wave function.

In natural units ($\hbar=c=1$), the MM operator for an S-wave multiquark state is
\begin{eqnarray}
    \hat{\mu}_z=\sum_i\hat{\mu}_{iz}=\sum_i\mu_i\hat{s}_z=\sum_ig\frac{q_i}{2m_i}\hat{s}_z,\label{eq:quark-magnetic-moment}
\end{eqnarray}
where $g$ is the $g$-factor, and $q_i$ and $m_i$ are the electric charge and mass of the $i$-th quark, respectively.

Using Eqs. \eqref{func:tetraquark} and \eqref{eq:quark-magnetic-moment}, the MM of an $S$-wave $QQ^\prime  \bar{q}\bar{q}^\prime$ tetraquark is given by 
\begin{eqnarray}
    \mu=\langle\Psi|\hat{\mu}_z|\Psi\rangle=\sum_{j i}\alpha_i\beta_j\langle\psi^{YIJ}_{QQ^\prime,j}|\hat{\mu}_z|\psi^{YIJ}_{QQ^\prime,i}\rangle,\label{eq:multiquark-magnetic-moment}
\end{eqnarray}
where the analytic expressions for $\langle\psi^{YIJ}_{QQ^\prime,j}|\hat{\mu}_z|\psi^{YIJ}_{QQ^\prime,i}\rangle$ are collected in the fifth column of \Cref{tab:magnetic-moment-QQqq}, \Cref{tab:magnetic-moment-QQns}, and \Cref{tab:magnetic-moment-bcqq} in the next section.

The radiative decay of an $S$-wave tetraquark $H$ to another tetraquark $H^\prime$ proceeds via magnetic dipole (M1) transition, with the MM operator $\hat{\mu}$ connecting the initial and final CMI eigenstates.
The corresponding partial width reads \cite{Wang:2022nqs}
\begin{eqnarray}\label{eq:radiative-decay-width}
	\Gamma&=&\frac{E^2_\gamma}{\pi}\frac{2}{2J_i+1}\sum_{J_{fz},J_{iz}}\frac{E_\gamma}{2}
	|\langle J_f,J_{fz}|\hat{\mu}_z|J_i,J_{iz}\rangle|^2,\nonumber\\
	&=&\alpha_{EM}\frac{E^3_\gamma}{m^2_p}\frac{1}{2J_i+1}\sum_{J_{fz},J_{iz}}\frac{|\langle J_f,J_{fz}|\hat{\mu}_z|J_i,J_{iz}\rangle|^2}{\mu^2_N},
\end{eqnarray}
where $\alpha_{EM}$ ($\approx1/137$) is the fine structure constant, $E_\gamma$ ($=\frac{M_H^2-M_{H^\prime}^2}{2M_H}$) denotes the emitted photon energy, $m_p$ is the mass of the proton, and $\mu_N=\frac{e}{2m_p}$ is the nuclear magneton where $e$ is the elementary charge. The transition amplitude is expanded in terms of the flavor-spin-color wave functions
\begin{eqnarray}
	&|J_i,J_{iz}\rangle=\sum_m\alpha_m|\psi^{YIJ_iJ_{iz}}_{QQ^\prime,m}\rangle, \quad
	|J_f,J_{fz}\rangle=\sum_n\beta_n|\psi^{YIJ_fJ_{fz}}_{QQ^\prime,n}\rangle,&\nonumber\\
	&\langle J_f,J_{fz}|\hat{\mu}_z|J_i,J_{iz}\rangle=\sum_{m,n}\alpha_m\beta_n\langle\psi^{IYJ_f J_{fz}}_{QQ^\prime,n}|\hat{\mu}_z|\Psi^{YIJ_iJ_{iz}}_{QQ^\prime,m}\rangle.&
\end{eqnarray}
We collect the squared amplitudes for all the $QQ^\prime  \bar{q}\bar{q}^\prime$ decay channels in \Cref{tab:ccnn-radiative-width,tab:QQns-radiative-width,tab:bcnn-radiative-width,tab:bcns-radiative-width,tab:bcss-radiative-width} in the next section.

\section{Numerical Results and Discussions}\label{sec3}

The evaluation of the radiative decay widths also requires the tetraquark masses as input. In Refs. \cite{Li:2023wug,Li:2023aui}, we presented a comprehensive analysis of the multiquark spectrum within the framework of a mass splitting model. The model provides a reasonable description of both the masses and the rearrangement decay patterns of the exotic states $T_{cc}(3875)^+$, $P_{\psi}^N(4457)^+$, $P_{\psi}^N(4440)^+$, $P_{\psi}^N(4337)^+$, $P_{\psi s}^\Lambda(4338)^0$, and $P_{\psi s}^\Lambda(4459)^0$.
Building on this success, we now employ the tetraquark masses and the effective coefficients $C_{ij}$ obtained there to study the magnetic moments and radiative decay widths of the S-wave doubly heavy tetraquark states.

\subsection{Parameters}

The mass splittings of conventional hadrons arising mainly from the chromomagnetic interaction can be used to extract the effective coupling coefficients $C_{ij}$ \cite{Wu:2016gas,Li:2023wug,Wu:2017weo,Li:2025fmf}. These coefficients depend on the quark flavors and the spatial wave functions in principle. 
For those between two quarks, we adopt values extracted from the mass splittings of conventional baryons, such as $N-\Delta$, $\Sigma-\Sigma^*$, $\Xi-\Xi^*$, $\Sigma_c-\Sigma_c^*$, $\Xi_c^\prime-\Xi_c^*$, $\Sigma_b-\Sigma^*_b$, and $\Xi_b^\prime-\Xi_b^*$.
The symmetry between matter and antimatter fields implies that the effective coupling coefficients between two antiquarks have the same values, e.g., $C_{\bar{c}\bar{s}}=C_{cs}$.
For a quark and an antiquark, the effective coupling is determined similarly, taking into account the meson mass splittings, such as $\pi-\rho$, $K-K^*$, $D-D^*$, $D_s-D_s^*$, $B-B^*$, $B_s-B_s^*$, $\eta_c-J/\psi$, and $\eta_b-\Upsilon$.
The numerical values of the effective coupling parameters we will use are collected in Table \ref{tab:parameters}.
\begin{table}[!h]
	\caption{Effective coupling coefficients $C_{ij}$ in units of MeV.}\label{tab:parameters}
	\begin{tabular}{c|cccccccc}
		\hline\hline
		$C_{ij}$&$b$&$c$&$s$&$n$&$\bar{b}$&$\bar{c}$&$\bar{s}$&$\bar{n}$\\\hline
		$b$&1.9&2.2&&&2.9&3.3&2.3&2.1\\
		$c$&&3.6&&&&5.3&6.7&6.6\\
		$n$&&&12.1&18.3&\\
		$s$&&&6.5\\
		\hline\hline
	\end{tabular}
\end{table}

The MMs for the $u$ and $d$ quarks are determined to be
\begin{eqnarray}
	&\mu_u=\frac{4\mu_p+\mu_n}{5}=1.85\mu_N,\quad\mu_d=\frac{\mu_p+4\mu_n}{5}=-0.97\mu_N&
\end{eqnarray}
where $\mu_p$ and $\mu_n$ are the MMs of the proton and neutron, respectively. Using the constituent quark masses $m_u=336$ MeV, $m_s=450$ MeV, and $m_c=1680$ MeV \cite{Wang:2022tib}, we extract the MMs of the $s$ and $c$ quarks as follows,
\begin{eqnarray}
	\mu_s=-\frac{m_u}{2m_s}\mu_u=-0.69\mu_N,\quad\mu_c=-\frac{m_u}{m_c}\mu_u=0.37\mu_N.
\end{eqnarray}
For the $b$ quark, its MM is determined to be
\begin{eqnarray}
	\mu_b=-\frac12\frac{m_c}{m_b}\mu_c=-\frac12\frac{m_{B_s^*}-m_{B_s}}{m_{D_s^*}-m_{D_s}}\mu_c=-0.06\mu_N.
\end{eqnarray}

\subsection{The magnetic moments of doubly heavy tetraquark states}

\subsubsection{The $QQ\bar{q}\bar{q}^\prime$ systems}

\begin{table}[!h]\centering
	\caption{CMI matrices, eigenvectors, and mass spectrum (in units of MeV) for the $QQ \bar{q}\bar{q}^\prime$ ($Q=b, c$; $q/q^\prime=n,s$) states. }\label{tab:mass-QQqq}
	\begin{tabular}{c|cccccccccc}\hline
		\hline\multicolumn{4}{c}{$cc\bar{n}\bar{n}^\prime$ system} \\\hline
		$I(J^{P})$ & $\langle H_{CMI} \rangle$ &Eigenvectors&Mass\\\hline
		$1(2^{+})$ &$\left(\begin{array}{c}93.3\end{array}\right)$&$\left(1\right)$&$\left(\begin{array}{c}4143.2\end{array}\right)$\\
		$1(1^{+})$ &$\left(\begin{array}{c}22.9\end{array}\right)$&$\left(1\right)$&$\left(\begin{array}{c}4072.8\end{array}\right)$\\
		$1(0^{+})$ &$\left(\begin{array}{cc}-12.3&129.3\\129.3&87.2\end{array}\right)$&$\left(\begin{array}{c}-0.82,0.57\\0.57,0.82\end{array}\right)$&$\left(\begin{array}{c}4225.9\\3948.8\end{array}\right)$\\
		$0(1^{+})$ &$\left(\begin{array}{cc}-137.1&-74.7\\-74.7&-10.4\end{array}\right)$&$\left(\begin{array}{c}0.42,-0.91\\-0.91,-0.42\end{array}\right)$&$\left(\begin{array}{c}4074.0\\3878.2\end{array}\right)$\\
		\hline\hline\multicolumn{4}{c}{$cc\bar{s}\bar{s}$ system} \\\hline
		$J^{P}$ & $\langle H_{CMI} \rangle$ &Eigenvectors  &Mass\\\hline
		$2^{+}$ &$\left(\begin{array}{c}62.4\end{array}\right)$&$\left(\begin{array}{c}1\end{array}\right)$&$\left(\begin{array}{c}4293.5\end{array}\right)$\\
		$1^{+}$ &$\left(\begin{array}{c}-9.1\end{array}\right)$&$\left(\begin{array}{c}1\end{array}\right)$&$\left(\begin{array}{c}4222.0\end{array}\right)$\\
		$0^{+}$ &$\left(\begin{array}{cc}-44.8&131.3\\131.3&40.0\end{array}\right)$&$\left(\begin{array}{c}-0.59,-0.81\\-0.81,0.59\end{array}\right)$&$\left(\begin{array}{c}4366.6\\4090.7\end{array}\right)$\\
		\hline\hline\multicolumn{4}{c}{$bb\bar{n}\bar{n}^\prime$ system} \\\hline
		$I(J^{P})$ & $\langle H_{CMI} \rangle$ &Eigenvectors &Mass\\\hline
		$1(2^{+})$ &$\left(\begin{array}{c}65.1\end{array}\right)$&$\left(\begin{array}{c}1\end{array}\right)$&$\left(\begin{array}{c}10795.3\end{array}\right)$\\
		$1(1^{+})$ &$\left(\begin{array}{c}42.7\end{array}\right)$&$\left(\begin{array}{c}1\end{array}\right)$&$\left(\begin{array}{c}10772.9\end{array}\right)$\\
		$1(0^{+})$ &$\left(\begin{array}{cc}31.5&41.2\\41.2&80.8\end{array}\right)$&$\left(\begin{array}{c}0.49,0.87\\-0.87,0.49\end{array}\right)$&$\left(\begin{array}{c}10834.4\\10738.4\end{array}\right)$\\
		$0(1^{+})$ &$\left(\begin{array}{cc}-141.3&-23.8\\-23.8&-16.8\end{array}\right)$&$\left(\begin{array}{c}0.18,-0.98\\-0.98,-0.18\end{array}\right)$&$\left(\begin{array}{c}10717.8\\10584.5\end{array}\right)$\\
		\hline\hline\multicolumn{4}{c}{$bb\bar{s}\bar{s}$ system} \\\hline
		$J^{P}$ & $\langle H_{CMI} \rangle$ &Eigenvectors&Mass\\\hline
		$2^{+}$ &$\left(\begin{array}{c}34.7\end{array}\right)$&$\left(\begin{array}{c}1\end{array}\right)$&$\left(\begin{array}{c}10946.1\end{array}\right)$\\
		$1^{+}$ &$\left(\begin{array}{c}10.1\end{array}\right)$&$\left(\begin{array}{c}1\end{array}\right)$&$\left(\begin{array}{c}10921.6\end{array}\right)$\\
		$0^{+}$ &$\left(\begin{array}{cc}-2.1&45.1\\45.1&33.6\end{array}\right)$&$\left(\begin{array}{c}0.56,0.83\\-0.83,0.56\end{array}\right)$&$\left(\begin{array}{c}10975.7\\10878.7\end{array}\right)$\\
        \hline\hline
\end{tabular}
\end{table}

The $QQ  \bar{q}\bar{q}^\prime$ ($Q=b, c$; $q/q^\prime=n,s$) states are classified into isospin triplets ($I=1$) and singlets ($I=0$). The three states labeled by $I_3$ in a triplet have different MMs, whereas there is only one MM for the singlet state. To get the MM values, one needs to know the CMI eigenstates of the $QQ \bar{q}\bar{q}^\prime$ states.

We adopt the CMI matrices and the corresponding mass spectra of the $QQ \bar{q}\bar{q}^\prime$ systems obtained in Ref. \cite{Li:2023wug} without modification. These results and the CMI eigenvectors for all the $QQ \bar{q}\bar{q}^\prime$ states are compiled in \Cref{tab:mass-QQqq}.
With the eigenvectors from the third column, one can write down the mass eigenstates of the S-wave $QQ \bar{q}\bar{q}^\prime$ tetraquark states using Eq.~\eqref{func:tetraquark}.
Subsequently, their MMs are computed via Eq.~\eqref{eq:multiquark-magnetic-moment}, with the results summarized in \Cref{tab:magnetic-moment-QQqq}.

\begin{table}[!h]
	\caption{Nonvanishing magnetic moments of the S-wave $QQ \bar{q}\bar{q}^\prime$ tetraquark states. The values of magnetic moments are in units of $\mu_N$.}\label{tab:magnetic-moment-QQqq}
	\begin{tabular}{c|c|c|c|c|ccccc}\hline\hline
		State&Isospin&$J^P$&Mass&$\langle\mu\rangle$&$I_3=1$&$I_3=0$&$I_3=-1$\\\hline
		\multirow{4}{*}{$cc\bar{n}\bar{n}^\prime$}
		&\multirow{2}{*}{$I=1$}
		&$2^+$&4143.2&$2\mu_c+\mu_{\bar{n}}+\mu_{\bar{n}^\prime}$&2.68&-0.14&-2.96\\\cline{3-8}
		&&$1^+$&4072.8&$\frac12(2\mu_c+\mu_{\bar{n}}+\mu_{\bar{n}^\prime})$&1.34&-0.07&-1.48\\\cline{2-8}
		&\multirow{2}{*}{$I=0$}
		&\multirow{2}*{$1^+$}&4074.0&\multirow{2}*{$2\alpha_1^2\mu_c+\alpha_2^2(\mu_{\bar{u}}+\mu_{\bar{d}})$}&&-0.59&\\
		&&&3878.2&&&0.45&\\\hline
		\multirow{2}{*}{$cc\bar{s}\bar{s}$}
		&\multirow{2}{*}{$I=0$}
		&$2^+$&4293.5&$2(\mu_c+\mu_{\bar{s}})$&&2.12\\\cline{3-8}
		&&$1^+$&4222.0&$\mu_c+\mu_{\bar{s}}$&&1.06\\\hline
		\multirow{4}{*}{$bb\bar{n}\bar{n}^\prime$}
		&\multirow{2}{*}{$I=1$}
		&$2^+$&10795.3&$2\mu_b+\mu_{\bar{n}}+\mu_{\bar{n}^\prime}$&1.82&-1.01&-3.83\\\cline{3-8}
		&&$1^+$&10772.9&$\frac12(2\mu_b+\mu_{\bar{n}}+\mu_{\bar{n}^\prime})$&0.91&-0.50&-1.92\\\cline{2-8}
		&\multirow{2}{*}{$I=0$}&\multirow{2}*{$1^+$}&10834.4&\multirow{2}*{$2\alpha_1^2\mu_b+\alpha_2^2(\mu_{\bar{u}}+\mu_{\bar{d}})$}&&-0.86&\\
		&&&10738.4&&&-0.15&\\\hline
		\multirow{2}{*}{$bb\bar{s}\bar{s}$}&\multirow{2}{*}{$I=0$}
		&$2^+$&10946.1&$2(\mu_b+\mu_{\bar{s}})$&&1.26\\\cline{3-8}
		&&$1^+$&10921.6&$\mu_b+\mu_{\bar{s}}$&&0.63\\\hline
	\end{tabular}
\end{table}

In the molecular picture, the MMs of the S-wave $cc\bar{n}\bar{n}^\prime$ states can be expressed in terms of the MMs of their constituent mesons.
For an $I(J^P)=1(2^+)$ state, its MM is
\begin{eqnarray}\label{eq:magnetic moment molecular12}
	\mu=\langle D^{*}D^{*}|\hat{\mu}_z|D^{*}D^{*}\rangle=2\mu_{D^*}=2\mu_c+\mu_{\bar{n}}+\mu_{\bar{n}^\prime}.
\end{eqnarray}
In the compact tetraquark picture, as shown in \Cref{tab:magnetic-moment-QQqq}, the MM of the $I(J^P)=1(2^+)$ $cc\bar{n}\bar{n}^\prime$   state has the same formal expression. This is because the highest-spin tetraquark has the unique spin configuration. 
Thus, it is impossible to distinguish tetraquark structures solely based on the MM measurement for the $I(J^P)=1(2^+)$ $cc\bar{n}\bar{n}^\prime$   case.

An $I(J^P)=1(1^+)$ $cc\bar{n}\bar{n}^\prime$ state in the molecular picture is a pure bound state of $D^*D$. Its wave function is
\begin{eqnarray}
	\begin{array}{lcl}
		|D^*D\rangle=\frac{1}{\sqrt{2}}\Big(D^{*+}_{11}D^+_{00}+D^+_{00}D^{*+}_{11}\Big)
		&&for\quad (I,I_3)=(1,1),\\ \\
		|D^*D\rangle=\frac{1}{2}\Big(D^{*+}_{11}D^0_{00}+D^{*0}_{11}D^+_{00}+D^+_{00}D^{*0}_{11}+D^0_{00}D^{*+}_{11}\Big)
		&&for\quad (I,I_3)=(1,0),\\ \\
		|D^*D\rangle=\frac{1}{\sqrt2}\Big(D^{*0}_{11}D^0_{00}+D^0_{00}D^{*0}_{11}\Big)
		&&for\quad (I,I_3)=(1,-1),
	\end{array}
\end{eqnarray}
where the subscript of a meson denotes its spin and the third component of the spin. 
The MM of this molecule can be expressed as
\begin{eqnarray}\label{eq:magnetic moment molecular11}
\mu&=&\langle D^{*}D|\hat{\mu}_z|D^{*}D\rangle=\mu_c+\frac12(\mu_{\bar{n}}+\mu_{\bar{n}^\prime}).
\end{eqnarray}
When $I_3=\pm1$, the condition $\mu_{\bar{n}}=\mu_{\bar{n}^\prime}$ simplifies this formula to
\begin{eqnarray}\label{eq:magnetic moment molecular11-2}
	\mu=\mu_c+\mu_{\bar{n}}.
\end{eqnarray}
By comparison, the magnetic moment of an $(I,I_3)=(1,\pm 1)$ compact $cc\bar{n}\bar{n}^\prime$ state using the same condition is
\begin{eqnarray}\label{eq:magnetic moment tetraquark11}
	\mu=\mu_c+\mu_{\bar{n}}.
\end{eqnarray}
From Eqs. \eqref{eq:magnetic moment molecular11-2} and \eqref{eq:magnetic moment tetraquark11}, the MMs of an $I(J^P)=1(1^+)$ state predicted by the molecular and compact tetraquark pictures are exactly identical. Therefore, measuring the MM cannot help distinguish between these two configurations for the $I(J^P)=1(1^+)$ $cc\bar{n}\bar{n}^\prime$ case, either.

For the $I(J^P)=0(1^+)$ case, the state wave functions are linear combinations of the $D^*D^*$ and $D^*D$ components in the molecular picture. The general form of a wave function reads $x|D^*D^*\rangle+y|D^{*}D\rangle$ with
\begin{eqnarray}
	\begin{array}{l}
		|D^*D^*\rangle=\frac{1}{2}(D^{*+}_{11}D^{*0}_{10}-D^{*0}_{11}D^{*+}_{10}
		-D^{*+}_{10}D^{*0}_{11}+D^{*0}_{10}D^{*+}_{11})\\
		\\
		|D^*D\rangle=\frac{1}{2}\Big(D^{*+}_{11}D^0_{00}-D^{*0}_{11}D^+_{00}-D^+_{00}D^{*0}_{11}+D^0_{00}D^{*+}_{11}\Big).
	\end{array}
\end{eqnarray}	
The expression for the MM of such a molecule is
\begin{eqnarray}\label{eq:molecular-magnetic-moment-DD0-mixing}
	\mu&=&x^{2}\langle D^*D^*|\hat{\mu}_z|D^*D^*\rangle+y^{2}\langle D^*D|\hat{\mu}_z|D^*D\rangle+2x y\langle DD^*|\hat{\mu}_z|D^*D^*\rangle\nonumber\\
	&=&\mu_c+\frac12(\mu_{\bar{u}}+\mu_{\bar{d}})+2x y[\mu_c-\frac12(\mu_{\bar{u}}+\mu_{\bar{d}})].
\end{eqnarray}
In the compact tetraquark picture, the MM is given by
\begin{eqnarray}\label{eq:tetraquark-magnetic-moment-DD0-mixing}
	\mu=2\alpha_1^2\mu_c+\alpha_2^2(\mu_{\bar{u}}+\mu_{\bar{d}}),
\end{eqnarray}
where $\alpha_1$ and $\alpha_2$ are the weight coefficients of the two spin-color configurations [see Eq. \eqref{func:tetraquark}].
According to Eqs. \eqref{eq:molecular-magnetic-moment-DD0-mixing} and \eqref{eq:tetraquark-magnetic-moment-DD0-mixing}, the contributions of quarks and antiquarks to the hadron MM are generally different between the molecular and compact cases. Specifically, the two expressions coincide when $\alpha_1^2=\alpha_2^2=1/2$ (the amplitudes of the two basis wave functions in the compact tetraquark are equal) and $x y=0$ (the molecule is a pure $D^*D^*$ or $D^*D$). Our tetraquark wave function does not satisfy this condition. If the observed $T_{cc}(3875)^+$ is a pure $D^*D$ molecule, its MM may serve as a key observable to distinguish between the molecular and compact configurations. Future measurements of the MM may therefore provide valuable insights into the internal structure of the exotic $T_{cc}(3875)^+$.

We take a look at the explicit values for the MMs of the compact $cc\bar{n}\bar{n}^\prime$   states shown in the last three columns of \Cref{tab:magnetic-moment-QQqq}.
In the $J^P=2^+$ case, the largest MM ($2.68\,\mu_N$) is for the state with $(I,I_3)=(1,1)$, and the smallest ($-2.96\,\mu_N$) is for $(I,I_3)=(1,-1)$ since the $\bar{d}$ has larger MM than the $\bar{u}$. For the $I(J^P)=0(1^+)$ states, the spin-color mixing yields MMs of $-0.59\,\mu_N$ and $0.45\,\mu_N$ for the higher and lower tetraquarks, respectively. 
If the $T_{cc}(3875)^+$ is the low-mass $I(J^P)=0(1^+)$ compact $cc\bar{n}\bar{n}^\prime$, its MM is about $0.45\,\mu_N$. The corresponding $\alpha_1^2\approx 0.83$, a value larger than $1/2$. The value of $0.45\,\mu_N$ for the molecule MM corresponds to $x y\approx 0.32$, whereas the MM is $-0.07\,\mu_N$ for a pure $I(J^P)=0(1^+)$ $D^*D$ molecule. We hope that future measurements of MM could shed light on its substructure. For comparison, we collect other theoretical preditions for the MM of this state in \Cref{tab:magnetic-moment-Tcc}. 

\begin{table}[h!]
\caption{Predictions for magnetic moment of the $T_{cc}(3875)^+$ state in the literature. The values are given in units of $\mu_N$.}\label{tab:magnetic-moment-Tcc}
\centering
\begin{tabular}{|c|c|c|c|}
\hline
\textbf{Journal} & \textbf{Method} & \textbf{Structure Assumption} &  \(\mu_N\)\\
\hline
This work&Mass Splitting Model&Compact tetraquark&0.45\\
\hline
Ref.~\cite{Zhang:2025wmr}&Chromomagnetic interaction&Compact tetraquark&0.65\\
\hline
Ref.~\cite{Lei:2023ttd} & Constituent Quark Model & Molecular S-wave & \(-0.09\) \\
\hline
 \multirow{2}{*}{Ref.~\cite{Azizi:2021aib}}& \multirow{2}{*}{QCD Light-cone Sum Rules}  & Compact diquark-antidiquark & \(0.66^{+0.34}_{-0.23}\) \\
\cline{3-4}
 &  & Molecular \(D^*D + DD^*\) & \(0.43^{+0.23}_{-0.22}\) \\
\hline
Ref.~\cite{Mutuk:2023oyz} & Diffusion Monte Carlo & Four-quark state & \(-0.28\) \\
\hline
Ref.~\cite{Wu:2022gie} & Constituent Quark Model + HADS & Compact diquark-antidiquark & \(0.73 \sim 0.76\) \\
\hline
\end{tabular}
\end{table}

The MMs of the $bb\bar{n}\bar{n}^\prime$   states can be obtained by replacing the charm quark with the bottom quark in the MM expressions for the $cc\bar{n}\bar{n}^\prime$   case. Because the coefficient in front of $\mu_c$ is positive and $\mu_b<0<\mu_c$, this replacement leads to smaller MMs of the $bb\bar{n}\bar{n}^\prime$  states than those of the $cc\bar{n}\bar{n}^\prime$ states, as shown in \Cref{tab:magnetic-moment-QQqq}.
The $J^P=2^+$ state with $(I,I_3)=(1,-1)$ in the $bb\bar{n}\bar{n}^\prime$ case has a MM of $-3.83$ $\mu_N$, which is smaller than the corresponding value of $-2.96$ $\mu_N$ in the charm case.
Similarly, the $J^P=1^+$ $bb\bar{n}\bar{n}^\prime$ state with $(I,I_3)=(1,0)$ has a MM of $-0.50$ $\mu_N$, which is smaller than $-0.07$ $\mu_N$ in the corresponding charm case.
For the $I(J^P)=0(1^+)$ states in the $bb\bar{n}\bar{n}^\prime$ system, their MMs are $-0.86$ $\mu_N$ and $-0.15$ $\mu_N$ for the higher and lower states, respectively, both of which are smaller than the corresponding $cc\bar{n}\bar{n}^\prime$ states.
The mixing between the $B^*B^*$ and $BB^*$ molecular configurations affects the MMs of the $I(J^P)=0(1^+)$ $bb\bar{n}\bar{n}^\prime$ states, a behavior analogous to that in the $cc\bar{n}\bar{n}^\prime$ case.
This implies that measuring the MMs of these $bb\bar{n}\bar{n}^\prime$ states could also help discriminate between molecular and compact tetraquark structures.

We now turn to the compact $cc\bar{s}\bar{s}$ and $bb\bar{s}\bar{s}$ tetraquark states. Their MMs, as shown in \Cref{tab:magnetic-moment-QQqq}, are determined entirely by the heavy quark and strange antiquark contributions.
The $2^+$ and $1^+$ states in both systems have positive MMs, with values of $2.12$ $\mu_N$ and $1.06$ $\mu_N$ for the $cc\bar{s}\bar{s}$ system, and $1.26$ $\mu_N$ and $0.63$ $\mu_N$ for the $bb\bar{s}\bar{s}$ system, respectively. 
Given the analogy with the $(I,I_3)=(1,1)$ $cc\bar{d}\bar{d}$ states, the MMs of the $J^P=2^+$ $D_s^*D_s^*$ molecule and the $J^P=1^+$ $D_s^*D_s$ molecule are identical to those of the corresponding compact tetraquark states,
\begin{align*}
&\mu_{2^+}=\langle D_s^{*}D_s^{*}|\hat{\mu}_z|D_s^{*}D_s^{*}\rangle=2\mu_{D_s^*}=2(\mu_c+\mu_{\bar{s}})=2.12\,\mu_N,
\\
&\mu_{1^+}=\langle D_s^{*}D_s|\hat{\mu}_z|D_s^{*}D_s\rangle=\mu_c+\mu_{\bar{s}}=1.06\,\mu_N.
\end{align*}
Similarly, the MMs of the $2^+$ and $1^+$ molecules in the $bb\bar{s}\bar{s}$ case are obtained as 1.26$\,\mu_N$ and $0.63\,\mu_N$, respectively.

\subsubsection{The $cc\bar{n}\bar{s}$, $bb\bar{n}\bar{s}$, and $bc\bar{n}\bar{s}$ systems}

\begin{table}[!h]\centering
	\caption{CMI  matrices, eigenvectors, and mass spectrum (in units of MeV) for the $bb\bar{n}\bar{s}$, $cc\bar{n}\bar{s}$, and $bc\bar{n}\bar{s}$ states. }\scriptsize\label{tab:mass-QQns}
	\begin{tabular}{c|cccccccccc}\hline
		\hline\hline\multicolumn{4}{c}{$cc\bar{n}\bar{s}$ system} \\\hline
		$J^{P}$ & $\langle H_{CMI} \rangle$ & Eigenvectors &Mass\\\hline
		$2^{+}$ &$\left(\begin{array}{c}77.1\end{array}\right)$&$\left(\begin{array}{c}1\end{array}\right)$&$\left(\begin{array}{c}4217.5\end{array}\right)$\\
		$1^{+}$ &$\left(\begin{array}{ccc}6.1&-0.4&-0.8\\-0.4&-87.5&-75.2\\-0.8&-75.2&-2.1\end{array}\right)$&$\left(\begin{array}{c}-0.01,-0.50,0.86\\1.00,-0.01,0.01\\-0.01,-0.86,-0.50\end{array}\right)$&$\left(\begin{array}{c}4182.2\\4146.6\\4009.2\end{array}\right)$\\
		$0^{+}$ &$\left(\begin{array}{cc}-29.3&130.3\\130.3&62.4\end{array}\right)$&$\left(\begin{array}{c}0.58,0.82\\-0.82,0.58\end{array}\right)$&$\left(\begin{array}{c}4295.1\\4018.8\end{array}\right)$\\
		\hline\hline\multicolumn{4}{c}{$bb\bar{n}\bar{s}$ system} \\\hline
		$J^{P}$ & $\langle H_{CMI} \rangle$ &Eigenvectors &Mass\\\hline
		$2^{+}$ &$\left(\begin{array}{c}49.1\end{array}\right)$&$\left(\begin{array}{c}1\end{array}\right)$&$\left(\begin{array}{c}10869.9\end{array}\right)$\\
		$1^{+}$ &$\left(\begin{array}{ccc}25.6&-0.8&-1.6\\-0.8&-91.7&-24.9\\-1.6&-24.9&-8.5\end{array}\right)$&$\left(\begin{array}{c}1.00,0.00,-0.05\\0.05,-0.27,0.96\\-0.01,-0.96,-0.27\end{array}\right)$&$\left(\begin{array}{c}10846.5\\10819.1\\10722.2\end{array}\right)$\\
		$0^{+}$ &$\left(\begin{array}{cc}13.9&43.1\\43.1&56.0\end{array}\right)$&$\left(\begin{array}{c}0.53,0.85\\-0.85,0.53\end{array}\right)$&$\left(\begin{array}{c}10903.8\\10807.8\end{array}\right)$\\
\hline\hline\multicolumn{4}{c}{$bc\bar{n}\bar{s}$ system} \\\hline
$J^{P}$ & $\langle H_{CMI} \rangle$ &Eigenvectors &Mass\\\hline
$2^{+}$ &$\left(\begin{array}{cc}39.9&0.3\\0.3&61.7\end{array}\right)$&$\left(\begin{array}{c}0.01,1.00\\-1.00,0.01\end{array}\right)$&$\left(\begin{array}{c}7542.4\\7520.6\end{array}\right)$\\
$1^{+}$ &$\left(\begin{array}{cccccc}-78.1&-0.3&-1.4&35.6&42.0&-1.2\\-0.3&14.5&35.6&-0.6&-1.2&16.8\\-1.4&35.6&45.5&0&-0.3&-50.1\\35.6&-0.6&0&-90.9&-50.1&-0.1\\42.0&-1.2&-0.3&-50.1&-7.3&0\\-1.2&16.8&-50.1&-0.1&0&14.7\end{array}\right)$&$\left(\begin{array}{c}0.01,-0.29,-0.82,0.00,0.01,0.50\\0.04,-0.79,-0.08,-0.03,0.11,-0.59\\0.26,0.10,0.02,-0.32,0.90,0.07\\-0.16,-0.52,0.55,-0.12,0.01,0.62\\0.75,-0.09,0.12,0.63,0.00,0.13\\0.58,0.00,0.01,-0.70,-0.42,0.01\end{array}\right)$&$\left(\begin{array}{c}7568.7\\7511.3\\7503.0\\7437.0\\7432.1\\7330.2\end{array}\right)$\\
$0^{+}$ &$\left(\begin{array}{cccc}-137.1&-0.6&0.6&86.7\\-0.6&-9.1&86.7&0.2\\0.6&86.7&57.2&0\\86.7&0.2&0&-114.4\end{array}\right)$&$\left(\begin{array}{c}0.00,-0.57,-0.82,0.00\\0.66,-0.01,0.01,0.75\\0.00,-0.82,0.57,-0.01\\-0.75,0.00,0.00,0.66\end{array}\right)$&$\left(\begin{array}{c}7597.6\\7442.4\\7411.9\\7267.5\end{array}\right)$\\
\hline\hline
\end{tabular}
\end{table}

The only isospin of compact tetraquark states in the $cc\bar{n}\bar{s}$, $bb\bar{n}\bar{s}$, and $bc\bar{n}\bar{s}$ systems is $I=\frac12$.
With the mass splitting model, we get their CMI matrices, eigenvectors, and mass spectra shown in \Cref{tab:mass-QQns}.
The MMs obtained using Eq.~\eqref{eq:multiquark-magnetic-moment} are summarized in \Cref{tab:magnetic-moment-QQns}.

\begin{table}[!h]
	\caption{Nonvanishing magnetic moments of the S-wave $cc\bar{n}\bar{s}$, $bb\bar{n}\bar{s}$, and $bc\bar{n}\bar{s}$ tetraquark states. The values of magnetic moments are in units of $\mu_N$.}\label{tab:magnetic-moment-QQns}
	\begin{tabular}{c|c|c|c|c|ccccccc}\hline\hline
		System&Isospin&$J^P$&Mass&$\langle\mu\rangle$&$I_3=\frac12$&$I_3=-\frac12$\\\hline
		\multirow{4}{*}{$cc\bar{n}\bar{s}$}
		&\multirow{4}{*}{$I=\frac12$}
		&$2^+$&4217.5&$2\mu_c+\mu_{\bar{n}}+\mu_{\bar{s}}$&2.40&-0.42\\\cline{3-7}
		&&\multirow{3}*{$1^+$}&4182.2&\multirow{3}*{\begin{tabular}{c}$\frac12\alpha_1^2(2\mu_{c}+\mu_{\bar{n}}+\mu_{\bar{s}})+2\alpha_2^2\mu_{c}+$\\$\alpha_3^2(\mu_{\bar{n}}+\mu_{\bar{s}})+\sqrt{2}\alpha_1\alpha_2(\mu_{\bar{n}}-\mu_{\bar{s}})$\end{tabular}}&1.43&-0.71\\
		&&&4146.6&&1.20&-0.17\\
		&&&4009.2&&0.98&0.24\\\hline
		\multirow{4}{*}{$bb\bar{n}\bar{s}$}
		&\multirow{4}{*}{$I=\frac12$}
		&$2^+$&10869.9&$2\mu_b+\mu_{\bar{n}}+\mu_{\bar{s}}$&1.54&-1.29\\\cline{3-7}
		&&\multirow{3}*{$1^+$}&10846.5&\multirow{3}*{\begin{tabular}{c}$\frac12\alpha_1^2(2\mu_{b}+\mu_{\bar{n}}+\mu_{\bar{s}})+2\alpha_2^2\mu_{b}+$\\$\alpha_3^2(\mu_{\bar{n}}+\mu_{\bar{s}})+\sqrt{2}\alpha_1\alpha_2(\mu_{\bar{n}}-\mu_{\bar{s}})$\end{tabular}}&0.77&-0.66\\
		&&&10819.1&&1.53&-1.04\\
		&&&10722.2&&0.01&-0.23\\\hline
		\multirow{8}{*}{$bc\bar{n}\bar{s}$}
		&\multirow{8}{*}{$I=\frac12$}
		&\multirow{2}{*}{$2^+$}&7542.4&\multirow{2}{*}{$\mu_b+\mu_c+\mu_{\bar{n}}+\mu_{\bar{s}}$}&1.97&-0.85\\
		&&&7520.6&&1.97&-0.85\\\cline{3-7}
		&&\multirow{6}*{$1^+$}&7568.7&\multirow{5}*{\begin{tabular}{c}$\frac12(\alpha_1^2+\alpha_2^2)(\mu_b+\mu_c+\mu_{\bar{n}}+\mu_{\bar{s}})+$\\$(\alpha_3^2+\alpha_4^2)(\mu_b+\mu_c)+$\\$(\alpha_5^2+\alpha_6^2)(\mu_{\bar{n}}+\mu_{\bar{s}})-$\\$\sqrt{2}(\alpha_1\alpha_5+\alpha_2\alpha_6)(\mu_b-\mu_c)-$\\$\sqrt{2}(\alpha_1\alpha_3+\alpha_2\alpha_4)(\mu_{\bar{s}}-\mu_{\bar{n}})$\end{tabular}}&0.61&-0.18\\
		&&&7511.3&&1.52&-0.47\\
		&&&7503.0&&1.61&-0.71\\
		&&&7437.0&&0.82&-0.58\\
		&&&7432.1&&0.72&-0.25\\
		&&&7330.2&&0.63&-0.37\\\hline\hline	
	\end{tabular}
\end{table}

We first focus on the double-charm case. In the molecular picture, the MMs of the S-wave $cc\bar{n}\bar{s}$ states arise from their constituent mesons.
For a $J^P=2^+$ state, its MM is expressed as
\begin{eqnarray}\label{eq:magnetic moment molecular22}
\mu=\langle D^{*}D_s^{*}|\hat{\mu}_z|D^{*}D_s^{*}\rangle=\mu_{D^*}+\mu_{D_s^*}=2\mu_c+\mu_{\bar{n}}+\mu_{\bar{s}},
\end{eqnarray}
which is identical to that in the compact tetraquark picture, as shown in \Cref{tab:magnetic-moment-QQns}.
Thus, measuring the MM of the $J^P=2^+$ $cc\bar{n}\bar{s}$ state cannot distinguish between the molecular and compact tetraquark structures.

The formula for MMs of $J^P=1^+$ compact $cc\bar{n}\bar{s}$ states is
\begin{eqnarray}\label{eq:magnetic moment QQns J=1}
	\mu=\frac12\alpha_1^2(2\mu_{c}+\mu_{\bar{n}}+\mu_{\bar{s}})+2\alpha_2^2\mu_{c}+\alpha_3^2(\mu_{\bar{n}}+\mu_{\bar{s}})+\sqrt{2}\alpha_1\alpha_2(\mu_{\bar{n}}-\mu_{\bar{s}}),
\end{eqnarray}
where $\alpha_i$ are the weight coefficients in the compact tetraquark wave functions. 
The $\alpha_1^2$, $\alpha_2^2$, and $\alpha_3^2$ terms in Eq. \eqref{eq:magnetic moment QQns J=1} correspond to the spin configuration contributions of the $\psi^{(\frac13\frac121)}_{cc,1}$, $\psi^{(\frac13\frac121)}_{cc,2}$, and $\psi^{(\frac13\frac121)}_{cc,3}$ basis states, respectively. The last $\alpha_1\alpha_2$ term originates from the spin coupling between the $\psi^{(\frac13\frac121)}_{cc,1}$ and $\psi^{(\frac13\frac121)}_{cc,2}$ states. They correspond to the $S_{\bar{n}\bar{s}}=1$ and $S_{\bar{n}\bar{s}}=0$ cases, respectively, where $S_{\bar{n}\bar{s}}$ denotes the spin of the light antidiquark.

There are two types of $J^P=1^+$ molecule states: one corresponds to the $D^*D_s^*$ state and the other arises from the mixing between the $D^*D_s$ and $DD_s^*$ channels, a consequence of the proximity in their thresholds. The wave function of the mixed state has the form $\alpha D^*D_s +\beta DD_s^*$ with $\alpha^2+\beta^2=1$. 
The MM of the former type molecule is given by
\begin{eqnarray}\label{eq:magnetic moment molecular12-1}
	\mu=\mu_c+\frac12(\mu_{\bar{n}}+\mu_{\bar{s}}),
\end{eqnarray}
while that of the latter type molecule is expressed as
\begin{eqnarray}\label{eq:magnetic moment molecular12-2}
	\mu=\alpha^2\langle D^{*}D_s|\hat{\mu}_z|D^{*}D_s\rangle+\beta^2\langle DD_s^{*}|\hat{\mu}_z|DD_s^{*}\rangle
	=\mu_c+\alpha^2\mu_{\bar{n}}+\beta^2\mu_{\bar{s}},
\end{eqnarray}
using the wave functions
\begin{align*}
&|D^*D^*_s\rangle=\frac{1}{\sqrt{2}}(D^{*}_{11}D^{*}_{s 10}-D^{*}_{10}D^{*}_{s 11}),\\
&|D^*D_s\rangle=D^*_{11}D_{s00},\qquad |DD^*_s\rangle=D_{00}D^*_{s 11}.
\end{align*}
If the $J^P=1^+$ $cc\bar{n}\bar{s}$ state is a pure $D^*D^*_s$ molecule or a mixed $DD_s^*/D^*D_s$ one, its MM differs significantly from that of the compact tetraquark state.
In the general case, the mixing between these two type molecules may occur \cite{Li:2012ss}. The wave function has the form $x|D^*D_s^*\rangle+(y+z)/2|D^*D_s+DD_s^* \rangle+(y-z)/2|D^*D_s-DD_s^* \rangle$ with $x^2+y^2+z^2=1$. 
The expression for the MM of such a molecule is
\begin{eqnarray}\label{eq:three-channel}
\mu&=&x^2\langle D^*D_s^*|\hat{\mu}_z|D^*D_s^*\rangle +y^2\langle D^{*}D_s|\hat{\mu}_z|D^{*}D_s\rangle+z^2\langle DD_s^{*}|\hat{\mu}_z|DD_s^{*}\rangle\nonumber\\
&&+2xy\langle D^{*}D_s^*|\hat{\mu}_z|D^*D_s\rangle+2xz\langle D^{*}D_s^*|\hat{\mu}_z|DD_s^{*}\rangle\nonumber\\
&=&[1+2x(y+z)]\mu_c+[\frac12x^2-2xy+y^2]\mu_{\bar{n}}+[\frac12x^2-2xz+z^2]\mu_{\bar{s}}.
\end{eqnarray}
From Eqs. \eqref{eq:magnetic moment QQns J=1} and \eqref{eq:three-channel}, the MMs of a $J^P=1^+$ $cc\bar{n}\bar{s}$ state differs markedly between the compact tetraquark model and the molecular model in the general case.

To examine whether the MM is helpful to reveal the substructure of an observed $J^P=1^+$ tetraquark, we consider two idealized possibilities. 
($i$) The state may be interpreted as a pure $D^*D_s^*$ molecule or a compact $cc\bar{n}\bar{s}$. For either $\alpha_1=0, \alpha_2^2=\alpha_3^2=1/2$ or $\alpha_1=1,\alpha_2=\alpha_3=0$, the expressions of Eqs. \eqref{eq:magnetic moment QQns J=1} and \eqref{eq:magnetic moment molecular12-1} are identical in form and hence the MM alone cannot resolve the tetraquark structures. 
Our proposed compact $cc\bar{n}\bar{s}$ state, with a mass around 4147 MeV, approximately satisifies the latter condition. However, since it lies above the $D^*D_s^*$ threshold, interpreting it as a $D^*D_s^*$ molecule is less compelling.
($ii$) The state may be interpreted as a mixed $DD_s^*/D^*D_s$ molecule or a compact $cc\bar{n}\bar{s}$. 
One finds that Eqs. \eqref{eq:magnetic moment QQns J=1} and \eqref{eq:magnetic moment molecular12-2} are equal when $\alpha_2^2=\alpha_3^2$, $\alpha^2= \frac12+\sqrt{2}\alpha_1\alpha_2$ and $\beta^2= \frac12-\sqrt{2}\alpha_1\alpha_2$. If $\alpha^2=\beta^2=1/2$ (the amplitudes of $DD_s^*$ and $D^*D_s$ components in the molecule are equal), one has $\alpha_1\alpha_2=0$. Two cases follow: (a) $\alpha_1=1$, $\alpha_2=\alpha_3=0$ (axial-axial type diquark-antidiquark state) and (b) $\alpha_1=0$, $\alpha_2^2=\alpha_3^2=1/2$ (axial-scalar/scalar-axial type diquark-antidiquark state). Our intermediate $J^P=1^+$ $cc\bar{n}\bar{s}$ state around 4147 MeV roughly corresponds to the first case, but it is much heavier than the $DD_s^*/D^*D_s$ molecule.    
Thus, the MM alone cannot be used to distinguish internal configurations. In the general case, one should combine the information from spectrum, decay properties, and measured MM to reveal the structure of a $J^P = 1^+$ exotic $cc\bar{n}\bar{s}$ state.

Turning to the double-bottom case, we get results shown in \Cref{tab:mass-QQns} and \Cref{tab:magnetic-moment-QQns}. The MM of a $J^P=2^+$ compact $bb\bar{n}\bar{s}$ is smaller than that of its charmed counterpart because of the smaller $\mu_b$ than $\mu_c$.
However, the MM of a $J^P=1^+$ $bb\bar{n}\bar{s}$ is probably larger than that of its corresponding charmed state, underscoring the interaction complexity within tetraquark systems. 

In the $bc\bar{n}\bar{s}$ case, tetraquark states contain four flavors of quarks. The number of states doubles compared with the $cc\bar{n}\bar{s}$ or $bb\bar{n}\bar{s}$ case. The relevant results are also collected in \Cref{tab:mass-QQns} and \Cref{tab:magnetic-moment-QQns}. 
Since the two $J^P=2^+$ compact tetraquarks have the same spin configuration, their MMs are identical and the expressions are also equal to those in the molecule model.
In contrast, the MMs of the $J^P=1^+$ $bc\bar{n}\bar{s}$ states are affected by the mixing among the six flavor-spin-color configurations $\psi_{bc,i}^{(\frac13\frac121)}$ ($i=1,\dots,6$) in the compact tetraquark picture, resulting in a complicated expression.
Meanwhile, if a molecular $J^P=1^+$ $bc\bar{n}\bar{s}$ exists, e.g. $\bar{B}^*D_s^*$, $\bar{B}^*D_s$, $\bar{B}D_s^*$, $\bar{B}_s^*D^*$, $\bar{B}_s^*D$, or $\bar{B}_sD^*$, its MM would be given by the sum of the MMs of its constituent mesons. 
When the molecule is generally a mixed state of these components, with the wave functions
\begin{eqnarray}
t|\bar{B}^*D_s^*\rangle+x|\bar{B}_s^*D^*\rangle+v|\bar{B}^*D_s\rangle+w|\bar{B}D_s^*\rangle
+y|\bar{B}_s^*D\rangle+z|\bar{B}_sD^*\rangle
\end{eqnarray}
and
\begin{eqnarray}
&|\bar{B}^*D_s^*\rangle=\frac{1}{\sqrt{2}}(\bar{B}^{*}_{11}D^{*}_{s 10}-\bar{B}^{*}_{10}D^{*}_{s 11}),\quad|\bar{B}^*D_s\rangle=\bar{B}^{*}_{11}D_{s00},\quad
|\bar{B}D_s^*\rangle=\bar{B}_{00}D^{*}_{s 11},&\nonumber\\
&|\bar{B}_s^*D^*\rangle=\frac{1}{\sqrt{2}}(\bar{B}^{*}_{s 11}D^{*}_{10}-\bar{B}^{*}_{s 10}D^{*}_{11}),\quad|\bar{B}_s^*D\rangle=\bar{B}^{*}_{s 11}D_{00},\quad
|\bar{B}_sD^*\rangle=\bar{B}_{s 00}D^{*}_{11},&\nonumber
\end{eqnarray}
the MM is then expressed as
\begin{eqnarray}
\mu &=& \mu_{b}(\frac{t^{2}}{2}+v^{2}+\frac{x^{2}}{2}+y^{2}-\sqrt{2}tw-\sqrt{2}xz)
+ \mu_{c}(\frac{t^{2}}{2}+w^{2}+\frac{x^{2}}{2}+z^{2}+\sqrt{2}tv+\sqrt{2}xy)+\nonumber\\
&&\mu_{\bar{n}}(\frac{t^{2}}{2}+v^{2}+\frac{x^{2}}{2}+z^{2}+\sqrt{2}tw-\sqrt{2}xy)
+\mu_{\bar{s}}(\frac{t^{2}}{2}+w^{2}+\frac{x^{2}}{2}+y^{2}-\sqrt{2}tv+\sqrt{2}xz).
\end{eqnarray}
One finds that the MMs of the compact and molecular $J^P=1^+$ $bc\bar{n}\bar{s}$ states differ, providing a means to distinguish between these two structures.

\subsubsection{The $bc \bar{q}\bar{q}^\prime$ systems}

With the mass splitting model, we get the CMI matrices, eigenvectors, and mass spectrum for the compact $bc \bar{q}\bar{q}^\prime$ ($q/q^\prime=n,s$) states in \Cref{tab:mass-QQpqq}. 
Note that the CMI matrices for the $bc\bar{n}\bar{n}^\prime$ system are slightly changed compared with those in Ref. \cite{Cheng:2020nho} because of the difference in effective coupling parameters. 
The calculated MMs using Eq.~\eqref{eq:multiquark-magnetic-moment} are summarized in \Cref{tab:magnetic-moment-bcqq}.

\begin{table}[!h]\centering
	\caption{CMI matrices, eigenvectors, and mass spectrum (in units of MeV) for the $bc \bar{q}\bar{q}^\prime$ ($q/q^\prime=n,s$) states. }\scriptsize\label{tab:mass-QQpqq}
	\begin{tabular}{c|cccccccccc}\hline
\hline\hline\multicolumn{4}{c}{$bc\bar{n}\bar{n}^\prime$ system} \\\hline
$J^{P}$ & $\langle H_{CM} \rangle$ &Eigenvectors &Mass\\\hline
$1(2^{+})$ &$\left(\begin{array}{c}77.9\end{array}\right)$&$\left(\begin{array}{c}1.00\end{array}\right)$&$\left(\begin{array}{c}7468.9\end{array}\right)$\\
$1(1^{+})$ &$\left(\begin{array}{ccc}31.5&36.0&17.0\\36.0&70.3&-49.2\\17.0&-49.2&31.2\end{array}\right)$&$\left(\begin{array}{c}-0.29,-0.84,0.46\\0.79,0.07,0.61\\0.54,-0.54,-0.64\end{array}\right)$&$\left(\begin{array}{c}7499.8\\7437.7\\7365.6\end{array}\right)$\\
$1(0^{+})$ &$\left(\begin{array}{cc}8.3&85.2\\85.2&82.0\end{array}\right)$&$\left(\begin{array}{c}0.55,0.84\\-0.84,0.55\end{array}\right)$&$\left(\begin{array}{c}7528.1\\7342.3\end{array}\right)$\\
$0(2^{+})$ &$\left(\begin{array}{c}30.7\end{array}\right)$&$\left(\begin{array}{c}1.00\end{array}\right)$&$\left(\begin{array}{c}7420.7\end{array}\right)$\\
$0(1^{+})$ &$\left(\begin{array}{ccc}-85.3&36.0&42.4\\36.0&-140.5&-49.2\\42.4&-49.2&-15.6\end{array}\right)$&$\left(\begin{array}{c}0.32,-0.22,0.92\\0.83,0.54,-0.16\\0.46,-0.81,-0.36\end{array}\right)$&$\left(\begin{array}{c}7401.2\\7319.8\\7208.7\end{array}\right)$\\
$0(0^{+})$ &$\left(\begin{array}{cc}-143.3&85.2\\85.2&-164.0\end{array}\right)$&$\left(\begin{array}{c}-0.75,-0.66\\-0.66,0.75\end{array}\right)$&$\left(\begin{array}{c}7322.3\\7150.5\end{array}\right)$\\
\hline\hline\multicolumn{4}{c}{$bc\bar{s}\bar{s}$ system} \\\hline
$J^{P}$ & $\langle H_{CM} \rangle$ &Eigenvectors &Mass\\\hline
$2^{+}$ &$\left(\begin{array}{c}47.2\end{array}\right)$&$\left(\begin{array}{c}1.00\end{array}\right)$&$\left(\begin{array}{c}7618.5\end{array}\right)$\\
$1^{+}$ &$\left(\begin{array}{ccc}-0.8&35.2&16.6\\35.2&23.1&-50.9\\16.6&-50.9&-0.3\end{array}\right)$&$\left(\begin{array}{c}-0.28,-0.80,0.53\\0.81,0.09,0.58\\0.51,-0.59,-0.63\end{array}\right)$&$\left(\begin{array}{c}7640.0\\7586.2\\7509.6\end{array}\right)$\\
$0^{+}$ &$\left(\begin{array}{cc}-24.8&88.2\\88.2&34.8\end{array}\right)$&$\left(\begin{array}{c}0.58,0.81\\-0.81,0.58\end{array}\right)$&$\left(\begin{array}{c}7669.3\\7483.2\end{array}\right)$\\
\hline
\end{tabular}
\end{table}

\begin{table}[!h]
	\caption{Nonvanishing magnetic moments of the S-wave $bc \bar{q}\bar{q}^\prime$ tetraquark states. The values of magnetic moments are in units of $\mu_N$.}\label{tab:magnetic-moment-bcqq}
	\begin{tabular}{c|c|c|c|c|ccccccc}\hline\hline
		System&Isospin&$J^P$&Mass&$\langle\mu\rangle$&$I_3=1$&$I_3=0$&$I_3=-1$\\\hline
		\multirow{8}{*}{$bc\bar{n}\bar{n}^\prime$}
		&\multirow{4}{*}{$I=1$}
		&$2^+$&7468.9&$\mu_b+\mu_c+\mu_{\bar{n}}+\mu_{\bar{n}}$&2.25&-0.57&-3.40\\\cline{3-8}
		&&\multirow{3}*{$1^+$}&7499.8&\multirow{3}*{\begin{tabular}{c}$\frac12\alpha_1^2(\mu_{b}+\mu_{c}+\mu_{\bar{n}}+\mu_{\bar{n}})+\alpha_2^2(\mu_{b}+\mu_{c})+$\\$\alpha_3^2(\mu_{\bar{n}}+\mu_{\bar{n}})+\sqrt{2}\alpha_1\alpha_3(\mu_c-\mu_b)$\end{tabular}}&0.65&-0.08&-0.80\\
		&&&7437.7&&1.72&-0.21&-2.14\\
		&&&7365.6&&1.01&-0.57&-2.16\\\cline{2-8}
		&\multirow{4}{*}{$I=0$}
		&$2^+$&7420.7&$\mu_b+\mu_c+\mu_{\bar{d}}+\mu_{\bar{u}}$&&-0.57&\\\cline{3-8}
		&&\multirow{3}*{$1^+$}&7401.2&\multirow{3}*{\begin{tabular}{c}$\frac12\alpha_1^2(\mu_{b}+\mu_{c}+\mu_{\bar{d}}+\mu_{\bar{u}})+\alpha_2^2(\mu_{b}+\mu_{c})+$\\$\alpha_3^2(\mu_{\bar{d}}+\mu_{\bar{u}})+\sqrt{2}\alpha_1\alpha_3(\mu_c-\mu_b)$\end{tabular}}&&-0.58\\
		&&&7319.8&&&-0.21\\
		&&&7208.7&&&-0.07\\\hline
		\multirow{4}{*}{$bc\bar{s}\bar{s}$}
		&\multirow{4}{*}{$I=0$}
		&$2^+$&7618.5&$\mu_b+\mu_c+2\mu_{\bar{s}}$&&1.69\\\cline{3-8}
		&&\multirow{3}*{$1^+$}&7640.0&\multirow{3}*{\begin{tabular}{c}$\frac12\alpha_1^2(\mu_b+\mu_{c}+2\mu_{\bar{s}})+\alpha_2^2(\mu_b+\mu_c)+$\\$2\alpha_3^2\mu_{\bar{s}}+\sqrt{2}\alpha_1\alpha_3(\mu_c-\mu_b)$\end{tabular}}&&0.56\\
		&&&7586.2&&&1.31\\
		&&&7509.6&&&0.67\\\hline
	\end{tabular}
\end{table}

Similar to the $QQ\bar{n}\bar{n}^\prime$ case, the $bc\bar{n}\bar{n}^\prime$ states are classified into isotriplets ($I=1$) and isosinglets ($I=0$). Since $b$ and $c$ are not identical quarks, the angular momenta $J=2$ and $J=1$ are both allowed for tetraquarks in each multiplet.
The triplet state with $I_z=0$ and the singlet state share the same wave function bases, and consequently, their MM formulas are identical.
For the isospin triplet case, the MM of the $2^{+}$ state is given by $\mu_b + \mu_c + \mu_{\bar{n}} +\mu_{\bar{n}^\prime}$, with calculated values of $2.25\,\mu_N$, $-0.57\,\mu_N$, and $-3.40\,\mu_N$ for $I_3 = 1, 0, -1$, respectively, showing a clear isospin-dependent progression.
The MMs of the three $1^{+}$ states are influenced by the spin coupling between different $bc$ diquark states, which is reflected by the cross term $\sqrt{2}\alpha_1\alpha_3(\mu_c-\mu_b)$, a feature similar to the $cc\bar{n}\bar{s}$ case (Eq. \eqref{eq:magnetic moment QQns J=1}).
The MMs for the three $I_3=1$ states are $0.65\,\mu_N$, $1.72\,\mu_N$, and $1.01\,\mu_N$, respectively, spanning a relatively wide range and reflecting the complexity of their internal configuration.
For the isospin singlet case, the $(I,J)=(0,2)$ state has the same MM as the $(I,J)=(1,2)$ state.
Due to differences in weight coefficients ($\alpha_i$), the $(I,I_z,J)=(0,0,1)$ states have distinct MMs from their $(I,I_z,J)=(1,0,1)$ counterparts.
Numerically, however, an $(I,I_z,J)=(0,0,1)$ state may occasionally have the same MM with an $(I,I_z,J)=(1,0,1)$ state. For a $1^+$ molecule with the general wave function $t|\bar{B}^{*0}D^{*0}\rangle+v|\bar{B}^{*0}D^{0}\rangle+w|\bar{B}^{0}D^{*0}\rangle+x|\bar{B}^{*-}D^{*+}\rangle+y|\bar{B}^{*-}D^{+}\rangle+z|\bar{B}^{-}D^{*+}\rangle$,
its MM is
\begin{eqnarray}
\mu &=&\mu_{b}(\frac{t^{2}}{2}+v^{2}+\frac{x^{2}}{2}+y^{2}-\sqrt{2}tw-\sqrt{2}xz)
+ \mu_{c}(\frac{t^{2}}{2}+w^{2}+\frac{x^{2}}{2}+z^{2}+\sqrt{2}tv+\sqrt{2}xy)+\nonumber\\
&&\mu_{\bar{d}}(\frac{t^{2}}{2}+v^{2}+\frac{x^{2}}{2}+z^{2}+\sqrt{2}tw-\sqrt{2}xy)
+\mu_{\bar{u}}(\frac{t^{2}}{2}+w^{2}+\frac{x^{2}}{2}+y^{2}-\sqrt{2}tv+\sqrt{2}xz).
\end{eqnarray}

In the $bc\bar{s}\bar{s}$ case, isospin is not relevant. The expressions can be obtained from the $bc\bar{n}\bar{n}^\prime$ case with the replacements $\mu_{\bar u}\to \mu_{\bar s}$ and $\mu_{\bar d}\to \mu_{\bar s}$. All the $bc\bar{s}\bar{s}$ MMs are numerically positive. 

\subsubsection{Discussions}

From the above discussions, unified expressions for the MMs of the $S$-wave compact $Q_1Q_2\bar{q}_1\bar{q}_2$ tetraquark states can be summarized as follows.
\begin{eqnarray}
\mu_{2^+}&=&\mu_{Q_1}+\mu_{Q_2}+\mu_{q_1}+\mu_{q_2},\\
\mu_{1^+}&=&\sum_c\Big[\frac12(\alpha_{1,1}^c)^2(\mu_{Q_1}+\mu_{Q_2}+\mu_{\bar{q}_1}+\mu_{\bar{q}_2})
+(\alpha_{1,0}^c)^2(\mu_{Q_1}+\mu_{Q_2})
+(\alpha_{0,1}^c)^2(\mu_{\bar{q}_1}+\mu_{\bar{q}_2})\nonumber\\
&&-\sqrt{2}(\alpha_{1,1}^c\alpha_{0,1}^c)(\mu_{Q_1}-\mu_{Q_2})
+\sqrt{2}(\alpha_{1,1}^c\alpha_{1,0}^c)(\mu_{\bar{q}_1}-\mu_{\bar{q}_2})\Big]\nonumber\\
&=&\sum_c\left(\begin{array}{ccc}
\alpha_{1,1}^c&\alpha_{1,0}^c&\alpha_{0,1}^c
\end{array}\right)
\left(\begin{array}{ccc}
\frac12(\mu_{Q_1}+\mu_{Q_2}+\mu_{\bar{q}_1}+\mu_{\bar{q}_2}) &\frac{1}{\sqrt2}(\mu_{\bar{q}_1}-\mu_{\bar{q}_2})&-\frac{1}{\sqrt2}(\mu_{Q_1}-\mu_{Q_2})  \\
\frac{1}{\sqrt2}(\mu_{\bar{q}_1}-\mu_{\bar{q}_2}) &(\mu_{Q_1}+\mu_{Q_2}) &0\\
-\frac{1}{\sqrt2}(\mu_{Q_1}-\mu_{Q_2}) & 0 & (\mu_{\bar{q}_1}+\mu_{\bar{q}_2})
\end{array}\right)
\left(\begin{array}{c}
\alpha_{1,1}^c\\\alpha_{1,0}^c\\\alpha_{0,1}^c
\end{array}\right).\label{muandM}
\end{eqnarray}
In the latter expression, the subscripts of the weight coefficient $\alpha_{a,b}^c$ denote the spins of the $Q_1Q_2$ diquark and the $\bar{q}_1\bar{q}_2$ antidiquark, respectively, while the superscript labels the color channel, $\bar{3}\otimes 3$ or $6\otimes\bar{6}$, of the diquark-antidiquark system. The summation runs over these two color channels. Obviously, the normalization condition is
\begin{eqnarray*}
\sum_c\Big[(\alpha_{1,1}^c)^2+(\alpha_{1,0}^c)^2+(\alpha_{0,1}^c)^2\Big]=1.
\end{eqnarray*}

Although the expression for the $1^+$ MM is correct, one has to notice that the weight coefficients may be zero because of the symmetry requirement. To explicitly incorporate the Pauli exclusion effect, we rewrite the expression to be
\begin{eqnarray}
\mu_{1^+}&=&X^T_1{\cal M}_1 X_1+X^T_2{\cal M}_2 X_2=
\left(\begin{array}{cc}
X_1^{T}&X_2^{T}
\end{array}\right)
\left(\begin{array}{cc}
{\cal M}_1&0\\0&{\cal M}_2
\end{array}\right)
\left(\begin{array}{cc}
X_1\\X_2
\end{array}\right)=\Psi^T\mathbb{M} \Psi,
\end{eqnarray}
where
$X_1=X^{\bar{3},3}=(\alpha_{1,1},\alpha_{1,0},\alpha_{0,1})^T_{\bar{3}\otimes 3}$,
$X_2=X^{6,\bar{6}}=(\alpha_{1,1},\alpha_{1,0},\alpha_{0,1})^T_{6\otimes\bar{6}}$, and
\begin{eqnarray}
{\cal M}_1&=&
\left(\begin{array}{ccc}
\mu_T\Big(\eta_q^A\Big) &\sqrt2\mu_{qq}\Big(\eta_q^A\eta_q^S\Big)&-\sqrt2\mu_{QQ}\Big(\eta_Q\eta_q^A\Big)\\
 &(\mu_{Q_1}+\mu_{Q_2})\Big(\eta_q^S\Big) &0\\
 & & (\mu_{\bar{q}_1}+\mu_{\bar{q}_2})\Big(\eta_Q\eta_q^A\Big)
\end{array}\right)
\nonumber\\
{\cal M}_2&=&\left(\begin{array}{ccc}
\mu_T\Big(\eta_Q\eta_q^S\Big) &\sqrt2\mu_{qq}\Big(\eta_Q\eta_q^S\eta_q^A\Big)&-\sqrt2\mu_{QQ}\Big(\eta_Q\eta_q^S\Big)  \\
 &(\mu_{Q_1}+\mu_{Q_2})\Big(\eta_Q\eta_q^A\Big) &0\\
 &  & (\mu_{\bar{q}_1}+\mu_{\bar{q}_2})\Big(\eta_q^S\Big)
\end{array}\right),
\end{eqnarray}
with $\mu_T=\frac12(\mu_{Q_1}+\mu_{\bar{q}_1}+\mu_{Q_2}+\mu_{\bar{q}_2})$,  $\mu_{QQ}=\frac12(\mu_{Q_1}-\mu_{Q_2})$, $\mu_{qq}=\frac12(\mu_{\bar{q}_1}-\mu_{\bar{q}_2})$. The lower-triangular portions of the symmetric ${\cal M}_{1}$ and ${\cal M}_{2}$ are not explicitly provided. The three introduced factors are defined as follows \cite{Luo:2017eub}: (i) When the two light quarks in flavor space are symmetric (antisymmetric), $\eta_q^S=0$ ($\eta_q^A=0$); (ii) When the two heavy quarks are identical, $\eta_Q=0$; (iii) The factor is 1 in other cases. Now, the matrices ${\cal M}_1$ and ${\cal M}_2$ are usually different and they operate differently in the $\bar{3}\otimes 3$ and $6\otimes\bar{6}$ color channels. They are magnetic operators in the spin-flavor space (but affected by color representation), and can be calculated with the basis wave functions in table \ref{tab:tetraquark-wave-function}. The contributions to $\mu_{1^+}$ from the two color channels may be separately obtained with the matrices. The diagonal elements of ${\cal M}_{1,2}$ represent the magnetic moments of the individual spin-spin configurations of the diquark-antidiquark system. The off-diagonal elements characterized by the differences $(\mu_{Q_1}-\mu_{Q_2})$ and $(\mu_{\bar{q}_1}-\mu_{\bar{q}_2})$ describe the transitions between configurations differing in the $(Q_1Q_2)$ spin state and in the $(\bar{q}_1\bar{q}_2)$ spin state, respectively. The vanishing off-diagonal elements of ${\cal M}_{1,2}$ reflect the single-particle nature of the magnetic operator which forbids simultaneous spin-flips of both the diquark and the antidiquark subsystems. The coefficients in the matrices ${\cal M}_{1,2}$ originate from the couplings of spin wave functions. We will call the matrices ${\cal M}_{1,2}$ and $\mathbb{M}$ the magnetic coupling ones which characterize the magnetic moments of the tetraquark system with $J^P=1^+$.

If one of the introduced factors is equal to 0, the magnetic coupling matrix ${\cal M}$ may be reduced to a $2\times2$ matrix or a number. When $Q_1\neq Q_2$ or $q_1\neq q_2$, the matrix ${\cal M}_{1,2}$ may have nonvanishing off-diagonal elements and the MM of the compact tetraquark can reflect the configuration mixing in the spin-flavor space. Diagonalizing a magnetic coupling matrix ${\cal M}$ yields its eigenvalues $\mathtt{m}_i$ and the corresponding eigenvectors $U_i$. The MM of a compact tetraquark can then be expressed as
\begin{eqnarray}
\mu_{1^+}=\sum_i\Big(p_i^{(1)}\mathtt{m}_i^{(1)}+p_i^{(2)}\mathtt{m}_i^{(2)}\Big),\quad p_i^{(1)}\equiv(U_i^{(1)\,T}X_1)^2,\quad p_i^{(2)}\equiv(U_i^{(1)\,T}X_2)^2.
\end{eqnarray}
The weights $p_i$ satisfying $\sum_i(p_i^{(1)}+p_i^{(2)})=1$ reflect the overall probability of the tetraquark state to be found in each eigen-configuration of ${\cal M}_{1,2}$. This decomposition of $\mu_{1^+}$ allows us to interpret the tetraquark MM as a weighted average over the eigenvalues of the magnetic coupling matrices. If ${\cal M}_1$ (or ${\cal M}_2$) is diagonal, its eigenvectors $U^{(1)}_i$ (or $U^{(2)}_i$) coincide with the basis vectors $(1,0,0)^T$, $(0,1,0)^T$, and $(0,0,1)^T$. In this case, $p_i^{(1)}=(X^{\bar{3},3}_{i})^2$ and $p_i^{(2)}=(X^{6,\bar{6}}_i)^2$.

In the general case, the tetraquark MM is strictly bounded by the extreme eigenvalues
\begin{eqnarray}
\mathtt{m}_{min}\leq \mu_{1^+}\leq\mathtt{m}_{max},
\end{eqnarray}
where $\mathtt{m}_{min}$ ($\mathtt{m}_{max}$) denotes the minimum (maximum) eigenvalue of the two magnetic coupling matrices or that of $\mathbb{M}$. This inequality provides a useful self-consistency check for the numerical results. If $\mu_{1^+}$ approaches either bound, the physical tetraquark is dominated by a single spin-flavor configuration. If $\mu_{1^+}$ falls in the intermediate region, multiple spin-flavor configurations would contribute to the tetraquark state.

For an $S$-wave doubly heavy molecule state, we only need to discuss the $J^P=1^+$ case. There are two types of quark combinations: $(Q_1\bar{q}_1)(Q_2\bar{q}_2)$ and $(Q_1\bar{q}_2)(Q_2\bar{q}_1)$. The wave function of a general meson-meson molecule has the form
\begin{eqnarray}
x_{1,1}(Q_1\bar{q}_1)_1(Q_2\bar{q}_2)_1+x_{1,0}(Q_1\bar{q}_1)_1(Q_2\bar{q}_2)_0+x_{0,1}(Q_1\bar{q}_1)_0(Q_2\bar{q}_2)_1\nonumber\\
+y_{1,1}(Q_1\bar{q}_2)_1(Q_2\bar{q}_1)_1+y_{1,0}(Q_1\bar{q}_2)_1(Q_2\bar{q}_1)_0+y_{0,1}(Q_1\bar{q}_2)_0(Q_2\bar{q}_1)_1\,.
\end{eqnarray}
Here $(Q\bar{q})_J$ just denotes the quark content of a heavy-quark meson with spin $J$. To account for the symmetry requirement explicitly, we modify the above wave function to be
\begin{align}
&\frac{t}{\sqrt2}[(Q_1\bar{q}_1)_1(Q_2\bar{q}_2)_1+(Q_1\bar{q}_2)_1(Q_2\bar{q}_1)_1]\Big(\eta_{Qq}^S\eta_q^Af\Big)
+\frac{v}{\sqrt2}[(Q_1\bar{q}_1)_1(Q_2\bar{q}_2)_1-(Q_1\bar{q}_2)_1(Q_2\bar{q}_1)_1]\Big(\eta_q^S\Big)
\nonumber\\
&+\frac{w}{2}\Big\{[(Q_1\bar{q}_1)_1(Q_2\bar{q}_2)_0+(Q_1\bar{q}_2)_1(Q_2\bar{q}_1)_0]
+[(Q_1\bar{q}_1)_0(Q_2\bar{q}_2)_1+(Q_1\bar{q}_2)_0(Q_2\bar{q}_1)_1]\Big\}\Big(\eta_q^Af\Big)
\nonumber\\
&+\frac{x}{2}\Big\{[(Q_1\bar{q}_1)_1(Q_2\bar{q}_2)_0+(Q_1\bar{q}_2)_1(Q_2\bar{q}_1)_0]
-[(Q_1\bar{q}_1)_0(Q_2\bar{q}_2)_1+(Q_1\bar{q}_2)_0(Q_2\bar{q}_1)_1]\Big\}\Big(\eta_Q\eta_q^Af\Big)
\nonumber\\
&+\frac{y}{2}\Big\{[(Q_1\bar{q}_1)_1(Q_2\bar{q}_2)_0-(Q_1\bar{q}_2)_1(Q_2\bar{q}_1)_0]
+[(Q_1\bar{q}_1)_0(Q_2\bar{q}_2)_1-(Q_1\bar{q}_2)_0(Q_2\bar{q}_1)_1]\Big\}\Big(\eta_Q\eta_q^S\Big)
\nonumber\\
&+\frac{z}{2}\Big\{[(Q_1\bar{q}_1)_1(Q_2\bar{q}_2)_0-(Q_1\bar{q}_2)_1(Q_2\bar{q}_1)_0]
-[(Q_1\bar{q}_1)_0(Q_2\bar{q}_2)_1-(Q_1\bar{q}_2)_0(Q_2\bar{q}_1)_1]\Big\}\eta_q^S
\end{align}
by recombining the terms and adding correction factors. In the wave function, the factor $f=1/\sqrt{2}$ (if $q_1=q_2$) or $f=1$ (if $q_1\neq q_2$). The factor $\eta_{Qq}^S=\eta_Q+\eta_q^S-\eta_Q\eta_q^S$ is 0 if both $Q_1Q_2$ and $q_1q_2$ are symmetric in flavor space, while it is 1 in other cases. It reflects the symmetry requirement for two identical bosons. The six coefficients satisfy the normalization condition $t^2+v^2+w^2+x^2+y^2+z^2=1$. One may similarly get
\begin{eqnarray}
\mu_{1^+}=\Psi^T\mathbb{M}\Psi
\end{eqnarray}
with $\Psi=(t,v,w,x,y,z)^T$ and
\begin{align}
{\mathbb M}=
\left(\begin{array}{cccccc}
\mu_T\Big(\eta_{Qq}^S\eta_Q\eta_q^A\Big) &0&-\mu_{QQ}\Big(\eta_{Qq}^S\eta_Q\eta_q^A\Big)&\mu_S\Big(\eta_{Qq}^S\eta_Q\eta_q^A\Big)&\mu_{qq}\Big(\eta_{Qq}^S\eta_Q\eta_q^S\eta_q^A\Big)&0  \\
 &\mu_T\Big(\eta_q^S\Big) &\mu_{qq}\Big(\eta_q^S\eta_q^A\Big)&0&-\mu_{QQ}\Big(\eta_Q\eta_q^S\Big)&\mu_S\Big(\eta_q^S\Big)\\
 &  & \mu_T\Big(\eta_q^A\Big)&\mu_{QQ}\Big(\eta_Q\eta_q^A\Big)&0&\mu_{qq}\Big(\eta_q^S\eta_q^A\Big)\\
&&&\mu_T\Big(\eta_Q\eta_q^A\Big)&\mu_{qq}\Big(\eta_Q\eta_q^S\eta_q^A\Big)&0\\
&&&&\mu_T\Big(\eta_Q\eta_q^S\Big)&\mu_{QQ}\Big(\eta_Q\eta_q^S\Big)\\
&&&&&\mu_T\Big(\eta_q^S\Big)
\end{array}\right)
\end{align}
with $\mu_S=\frac12(\mu_{Q_1}+\mu_{Q_2}-\mu_{\bar{q}_1}-\mu_{\bar{q}_2})$.

Different from the compact tetraquark case, the expression for the molecule state relies on one magnetic coupling matrix. The reason is that a hadronic molecule contains both color-$\bar{3}$ and color-$6$ $Q_1Q_2$ configurations. Let the eigenvalue equation of this matrix be
\begin{eqnarray}
\mathbb{M}U_i=\mathtt{m}_iU_i.
\end{eqnarray}
The MM of the molecular state can be rewritten as
\begin{eqnarray}
\mu_{1^+}=\sum_i\left(p_i\mathtt{m}_i\right),
\quad p_i\equiv(U_i^{T}\Psi)^2
\end{eqnarray}
where $\sum_{i}p_i=1$. One can also get an inequality
\begin{eqnarray}
\mathtt{m}_{min}\leq \mu_{1^+}\leq\mathtt{m}_{max},
\end{eqnarray}
where $\mathtt{m}_{min}$ ($\mathtt{m}_{max}$) denotes the minimum (maximum) eigenvalue of the magnetic coupling matrix $\mathbb{M}$. Apparently, checking whether the measured MM of a tetraquark state is in the range $\mathtt{m}_{min}^{compact}\leq \mu_{1^+}\leq\mathtt{m}_{max}^{compact}$ or $\mathtt{m}_{min}^{molecule}\leq \mu_{1^+}\leq\mathtt{m}_{max}^{molecule}$ may provide us the information of its internal structure. However, in all the $Q_1Q_2\bar{q}_1\bar{q}_2$ cases, the obtained $\mathtt{m}_{min,max}$ within these two pictures are identical, if all the contribution channels are included.

Since the tetraquark is a mixed state of different diquark-antidiquark or meson-meson configurations, the weight coefficients depend on the adopted models. The resulting $\mu_{1^+}$ values exhibit model-dependent discrepancies, but they always falls within the bounded range $[\mathtt{m}_{min},\mathtt{m}_{max}]$. When the tetraquark is a state of single configuration, one has $\mu_{1^+}=\mathtt{m}_{min}$ or $\mathtt{m}_{max}$. For example, for an $I(J^P)=0(1^+)$ compact $cc\bar{u}\bar{d}$ state, we have
\begin{eqnarray}
\mathtt{m}_{min}^{compact}=-0.88\,\mu_N,\qquad \mathtt{m}_{max}^{compact}=0.74\,\mu_N.
\end{eqnarray}
The MM of the state corresponding to $T_{cc}(3875)^+$ in our model is $0.45\,\mu_N$. It is in the range $[-0.88\,\mu_N,0.74\,\mu_N]$ and is closer to the upper bound which is the MM of the $[(cc)_{\bar{3}_c}^1(ud)_{3_c}]_{1_c}^1$ state. In the molecule picture, the values of $\mathtt{m}_{min}$ and $\mathtt{m}_{max}$ are the same as the compact case if the $D^*D^*$ component is included. If the $T_{cc}(3875)^+$ is a pure $DD^*$ molecule, only one element of $\mathbb{M}$ is nonvanishing and we have
\begin{eqnarray}
\mu_{1^+}=\mathtt{m}_{min}^{molecule}=\mathtt{m}_{max}^{molecule}=-0.07\,\mu_N.
\end{eqnarray}

In Table \ref{minmaxMM}, we collect the minimum and maximum MMs for various $1^+$ doubly heavy tetraquark states. Since all the contribution channels are considered for a state, the ranges in the compact and molecule pictures are identical. The $\mathtt{m}_{min}=\mathtt{m}_{max}$ cases corresponding to $\eta_Q=\eta_q^S=0$ reflect the fact that only one tetraquark state exists. The coupling matrices for the $bc\bar{n}\bar{n}^\prime$ cases $I=1$ ($\eta_Q=\eta_q^S=0$) and $I=0$ ($\eta_Q=\eta_q^A=0$) are the same, which result in the same range $[\mathtt{m}_{min},\mathtt{m}_{max}]$. As a cross check, one may confirm that our numerical MMs do fall within the range $[\mathtt{m}_{min},\mathtt{m}_{max}]$.

\begin{table}[!h]
\caption{Range of magnetic moments $[\mu_{min},\mu_{max}]$ for the S-wave $J^P=1^+$ $QQ^\prime\bar{q}\bar{q}^\prime$ tetraquark states in both compact and molecule pictures in units of $\mu_N$.}\label{minmaxMM} 
\begin{tabular}{c|c|ccc}\hline\hline
System&Isospin&$I_3=1$&$I_3=0$&$I_3=-1$\\\hline
		
$cc\bar{n}\bar{n}^\prime$&$I=1$	&$[1.34,1.34]$&$[-0.07,-0.07]$&$[-1.48,-1.48]$\\
&$I=0$&&$[-0.88,0.74]$&\\\hline

$cc\bar{s}\bar{s}$&$I=0$&&$[1.06,1.06]$&\\\hline

$bb\bar{n}\bar{n}^\prime$&$I=1$	&$[0.91,0.91]$&$[-0.50,-0.50]$&$[-1.91,-1.91]$\\
		&$I=0$&&$[-0.88,-0.12]$&\\\hline
		
$bb\bar{s}\bar{s}$&$I=0$&&$[0.63,0.63]$&\\\hline

$bc\bar{n}\bar{n}^\prime$&$I=1$	&$[0.31,2.04]$&$[-1.01,0.31]$&$[-3.75,0.31]$\\
&$I=0$&&$[-1.01,0.31]$&\\\hline

$bc\bar{s}\bar{s}$&$I=0$&&$[0.31,1.52]$&\\\hline\hline

	\end{tabular}

\begin{tabular}{c|c|cc}\hline\hline
	System&Isospin&$I_3=\frac12$&$I_3=-\frac12$\\\hline
	$cc\bar{n}\bar{s}$&$I=\frac12$	&$[0.67,1.66]$&$[-1.59,2.12]$	\\
	$bb\bar{n}\bar{s}$&$I=\frac12$	&$[-0.16,1.66]$&$[-2.19,1.43]$	\\
	$bc\bar{n}\bar{s}$&$I=\frac12$	&$[0.25,1.78]$&$[-1.96,1.79]$	
\\\hline
\end{tabular}
\end{table}

\subsection{The radiative decays of doubly heavy tetraquark states}

To reveal the intrinsic connections among different $QQ^\prime \bar{q}\bar{q}^\prime$ states, we further investigate the electromagnetic radiation properties of compact tetraquark states.
This study will provide a new perspective for distinguishing the internal structures of such states.

\subsubsection{The $QQ \bar{q}\bar{q}^\prime$   systems}

\begin{table}[!h]
	\caption{Radiative transitions of $QQ\bar{q}\bar{q}^\prime$ states in units of keV. A tetraquark is denoted as $T^{(I,J)}(Mass)$. In the third column, $\alpha_i$ and $\beta_i$ represent the weight coefficients of the spin-color configurations in the initial and final states, respectively.}\label{tab:ccnn-radiative-width}
	\begin{tabular}{c|c|c|cccccccc}\hline\hline
		\multirow{2}{*}{System}&\multirow{2}{*}{radiative decay channel}&\multicolumn{4}{c}{$\Gamma$ (keV)}\\\cline{3-6}
		&&$\langle\mu\rangle^2$&$I_3=1$&$I_3=0$&$I_3=-1$\\\cline{1-6}
		\multirow{12}{*}{$cc\bar{n}\bar{n}^\prime$}
		&$T_{cc}^{(1,2)}(4143)\rightarrow T_{cc}^{(1,1)}(4073)\gamma$&$\frac{5}{6}(-2\mu_c+\mu_{\bar{n}}+\mu_{\bar{n}^\prime})^2$&0.68&1.23&9.28\\\cline{3-6}
		&$T_{cc}^{(1,2)}(4143)\rightarrow T_{cc}^{(0,1)}(4074)\gamma$&\multirow{2}{*}{$\frac53\beta^2_1(\mu_{\bar{d}}-\mu_{\bar{u}})^2$}&&1.26\\
		&$T_{cc}^{(1,2)}(4143)\rightarrow T_{cc}^{(0,1)}(3878)\gamma$& &&306.37\\\cline{2-6}
		&$T_{cc}^{(1,1)}(4073)\rightarrow T_{cc}^{(1,0)}(3949)\gamma$&$\frac23\beta_1^2(-2\mu_c+\mu_{\bar{n}}+\mu_{\bar{n}^\prime})^2$&3.30&5.98&45.02\\\cline{3-3}
		&$T_{cc}^{(1,1)}(4073)\rightarrow T_{cc}^{(0,1)}(3878)\gamma$&$\beta_1^2(\mu_{\bar{d}}-\mu_{\bar{u}})^2$&&124.39\\\cline{2-6}
		&$T_{cc}^{(0,1)}(4074)\rightarrow T_{cc}^{(1,1)}(4073)\gamma$&$\alpha_1^2(\mu_{\bar{d}}-\mu_{\bar{u}})^2$&&$6.70\times 10^{-6}$&&\\\cline{3-3}
		&$T_{cc}^{(0,1)}(4074)\rightarrow T_{cc}^{(0,1)}(3878)\gamma$&$2(2\alpha_1\beta_1\mu_c+\alpha_2\beta_2(\mu_{\bar{d}}+\mu_{\bar{u}}))^2$&&14.72\\\cline{3-3}
		&$T_{cc}^{(0,1)}(4074)\rightarrow T_{cc}^{(1,0)}(3949)\gamma$& {$\frac19(\sqrt{3}\alpha_1\beta_1-3\alpha_2\beta_2)^2(\mu_{\bar{d}}-\mu_{\bar{u}})^2$}&& {4.07}&&\\\cline{2-6}
		&$T_{cc}^{(1,0)}(4226)\rightarrow T_{cc}^{(1,1)}(4073)\gamma$&$\frac23\alpha_1^2(-2\mu_c+\mu_{\bar{n}}+\mu_{\bar{n}^\prime})^2$&8.73&15.80&118.87\\\cline{3-6}
		&$T_{cc}^{(1,0)}(4226)\rightarrow T_{cc}^{(0,1)}(4074)\gamma$&\multirow{3}{*}{ {$\frac19(\sqrt{3}\alpha_1\beta_1-3\alpha_2\beta_2)^2(\mu_{\bar{d}}-\mu_{\bar{u}})^2$}}&& {172.04}\\
		&$T_{cc}^{(1,0)}(4226)\rightarrow T_{cc}^{(0,1)}(3878)\gamma$&&& {6.07}\\
		&$T_{cc}^{(1,0)}(3949)\rightarrow T_{cc}^{(0,1)}(3878)\gamma$&&& {10.16}\\
		\hline
		\multirow{12}{*}{$bb\bar{n}\bar{n}^\prime$}
		&$T_{bb}^{(1,2)}(10795)\rightarrow T_{bb}^{(1,1)}(10773)\gamma$&$\frac{5}{6}(-2\mu_b+\mu_{\bar{n}}+\mu_{\bar{n}^\prime})^2$&0.07&0.01&0.20\\\cline{3-6}
		&$T_{bb}^{(1,2)}(10795)\rightarrow T_{bb}^{(0,1)}(10718)\gamma$&\multirow{2}{*}{$\frac53\beta^2_1(\mu_{\bar{d}}-\mu_{\bar{u}})^2$}&& {0.33}\\
		&$T_{bb}^{(1,2)}(10795)\rightarrow T_{bb}^{(0,1)}(10585)\gamma$&& & {193.89}\\\cline{2-6}
		&$T_{bb}^{(1,1)}(10773)\rightarrow T_{bb}^{(1,0)}(10738)\gamma$&$\frac23\beta_1^2(-2\mu_b+\mu_{\bar{n}}+\mu_{\bar{n}^\prime})^2$&0.24&0.03&0.73\\\cline{3-6}
		&$T_{bb}^{(1,1)}(10773)\rightarrow T_{bb}^{(0,1)}(10718)\gamma$&\multirow{2}{*}{$\beta_1^2(\mu_{\bar{u}}-\mu_{\bar{d}})^2$}&&0.12\\
		&$T_{bb}^{(1,1)}(10773)\rightarrow T_{bb}^{(0,1)}(10585)\gamma$&&&138.84\\\cline{2-6}
		&$T_{bb}^{(0,1)}(10718)\rightarrow T_{bb}^{(0,1)}(10585)\gamma$&$2(2\alpha_1\beta_1\mu_b+\alpha_2\beta_2(\mu_{\bar{d}}+\mu_{\bar{u}}))^2$&&0.23\\\cline{2-6}
		&$T_{bb}^{(1,0)}(10834)\rightarrow T_{bb}^{(1,1)}(10773)\gamma$&$\frac23\alpha_1^2(-2\mu_b+\mu_{\bar{n}}+\mu_{\bar{n}^\prime})^2$&1.33&0.18&3.96\\\cline{3-6}
		&$T_{bb}^{(1,0)}(10834)\rightarrow T_{bb}^{(0,1)}(10718)\gamma$&\multirow{4}{*}{ {$\frac19(\sqrt{3}\alpha_1\beta_1-3\alpha_2\beta_2)^2(\mu_{\bar{d}}-\mu_{\bar{u}})^2$}}&& {84.90}\\
		&$T_{bb}^{(1,0)}(10834)\rightarrow T_{bb}^{(0,1)}(10585)\gamma$&&&  {14.87}\\
		&$T_{bb}^{(1,0)}(10738)\rightarrow T_{bb}^{(0,1)}(10718)\gamma$&&&  {0.09}\\
		&$T_{bb}^{(1,0)}(10738)\rightarrow T_{bb}^{(0,1)}(10585)\gamma$&&&  {80.27}&&\\
		\hline
		\multirow{3}{*}{$cc\bar{s}\bar{s}$}
		&$T_{cc}^{(0,2)}(4294)\rightarrow T_{cc}^{(0,1)}(4222)\gamma$&$\frac{10}{3}(\mu_c-\mu_{\bar{s}})^2$&&0.20\\\cline{3-3}
		&$T_{cc}^{(0,1)}(4222)\rightarrow T_{cc}^{(0,0)}(4091)\gamma$&$\frac83\beta_1^2(\mu_c-\mu_{\bar{s}})^2$&&1.07\\\cline{3-3}
		&$T_{cc}^{(0,0)}(4367)\rightarrow T_{cc}^{(0,1)}(4222)\gamma$&$\frac83\alpha_1^2(\mu_c-\mu_{\bar{s}})^2$&& {2.27}\\
		\hline
		\multirow{3}{*}{$bb\bar{s}\bar{s}$}
		&$T_{bb}^{(0,2)}(10946)\rightarrow T_{bb}^{(0,1)}(10922)\gamma$&$\frac{10}{3}(\mu_b-\mu_{\bar{s}})^2$&&0.05\\\cline{3-3}
		&$T_{bb}^{(0,1)}(10922)\rightarrow T_{bb}^{(0,0)}(10879)\gamma$&$\frac83\beta_1^2(\mu_b-\mu_{\bar{s}})^2$&&0.23\\\cline{3-3}
		&$T^{(0,0)}(10976)\rightarrow T_{bb}^{(0,1)}(10922)\gamma$&$\frac83\alpha_1^2(\mu_b-\mu_{\bar{s}})^2$&&0.62\\
		\hline\hline
	\end{tabular}	
\end{table}

The radiative decays of the $QQ \bar{q}\bar{q}^\prime$ states can be studied by using  Eq.~\eqref{eq:radiative-decay-width}, with the results presented in Table \ref{tab:ccnn-radiative-width}.
The amplitudes in the third column of Table \ref{tab:ccnn-radiative-width} contain the spin-color structure weights $\alpha_i$ or $\beta_i$, which indicates that the radiative transition between tetraquarks also encodes information about their inner structures. Obviously, the third component of isospin, $I_3$, affects the results.
Within an isospin triplet in the $QQ\bar{n}\bar{n}^\prime$ case, the ratio of the widths for $I_3=-1$ and $I_3=1$ states can be expressed by a unified formula,
\begin{eqnarray}\label{eq:width-ratio-QQnn}
	R=\Big(\frac{1-\mu_{\bar{u}}/\mu_Q}{1-\mu_{\bar{d}}/\mu_Q}\Big)^2.
\end{eqnarray}
Specifically, this ratio is approximately $13.60$ for the doubly charmed tetraquark system and about $3.00$ for the doubly bottom system.

In the $cc\bar{n}\bar{n}^\prime$ system, the amplitudes for channels involving $(I,J)=(0,1)$ states depend only on the MMs of the light antiquarks, except for the $T_{cc}^{(0,1)}(4074)\rightarrow T_{cc}^{(0,1)}(3878)\gamma$ channel whose amplitude also involves the MM of charm quark.
Such decay channels could be utilized to investigate in detail the structures of double-charm tetraquark states.
For example, the ratio of the width of $T_{cc}^{(1,2)}(4143)\rightarrow T_{cc}^{(0,1)}(4074)\gamma$ to the width of $T_{cc}^{(1,2)}(4143)\rightarrow T_{cc}^{(0,1)}(3878)\gamma$ can be used to measure the ratio of the weights of the color-triplet structure in $T_{cc}^{(0,1)}(4074)$ and $T_{cc}^{(0,1)}(3878)$.
For the radiative decay channel with the largest width, $T_{cc}^{(1,2)}(4143)\rightarrow T_{cc}^{(0,1)}(3878)\gamma$, the amplitude depends exclusively on the color-triplet structure of the final state $T_{cc}^{(0,1)}(3878)$, which is characterized by the spin-color structure weight $\beta_1 = -0.91$. During the decay, the two light antiquarks undergo a spin-flip transition from the $S_{\bar{n}\bar{n}^\prime}=1$ state to the $S_{\bar{n}\bar{n}^\prime}=0$ state. Owing to the large weight coefficient $\beta_1$ and the largest phase space, this decay channel yields a substantial width of $306.37$ keV.
The radiative decay channel with the smallest width, $T_{cc}^{(0,1)}(4074)\rightarrow T_{cc}^{(1,1)}(4073)\gamma$, also receives a contribution from the color-triplet structure of the initial $T_{cc}^{(0,1)}(4074)$ state. However, the associated weight $\alpha_1$ is only $0.42$, which is considerably smaller than the weight $\beta_1$ of the final state $T_{cc}^{(0,1)}(3878)$ in the decay channel $T_{cc}^{(1,2)}(4143)\rightarrow T_{cc}^{(0,1)}(3878)\gamma$. Moreover, the phase space for this decay is quite limited. These two factors together lead to an extremely small radiative decay width of $6.70\times 10^{-6}$ keV.
It is worth noting that if the observed $T_{cc}(3875)^+$ is identified as the predicted $T_{cc}^{(0,1)}(3878)$ compact state, five radiative decay channels from higher $cc\bar{n}\bar{n}^\prime$ states would open up, $T_{cc}^{(1,2)}(4143)$, $T_{cc}^{(1,1)}(4073)$, $T_{cc}^{(0,1)}(4074)$, $T_{cc}^{(1,0)}(4226)$, and $T_{cc}^{(1,0)}(3949)$.
Their predicted widths span from $6.07$ keV to $306.37$ keV.

Comparison of the radiative decay amplitudes for the $bb\bar{n}\bar{n}^\prime$ and $cc\bar{n}\bar{n}^\prime$ systems reveals a nearly identical decay pattern in form, confirming the heavy quark symmetry in the electromagnetic transitions.
In the $bb\bar{n}\bar{n}^\prime$ case, the broadest radiative decay channel is $T_{bb}^{(1,2)}(10795)\rightarrow T_{bb}^{(0,1)}(10585)\gamma$, with a width of $193.89$ keV.
In the cases of $cc\bar{s}\bar{s}$ and $bb\bar{s}\bar{s}$, there are three radiative decay channels for each system.
The radiative decays between the $J=1$ and $J=0$ states are only influenced by the color-triplet structure of the $J=0$ state.
The values of the radiative decay widths in the $cc\bar{s}\bar{s}$ system are in the range of $0.20$ keV to $2.27$ keV, while those in the $bb\bar{s}\bar{s}$ system range from $0.05$ keV to $0.62$ keV.

\subsubsection{The $QQ\bar{n}\bar{s}$ systems}

\begin{table}[htbp]
	\caption{Radiative transitions of $QQ\bar{n}\bar{s}$ states in units of keV. A tetraquark is denoted as $T^{(I,J)}(Mass)$. In the third column, $\alpha_i$ and $\beta_i$ represent the weight coefficients of the spin-color configurations in the initial and final states, respectively.}\label{tab:QQns-radiative-width}
	\begin{tabular}{c|c|c|ccccccc}\hline\hline
		\multirow{2}{*}{System}&\multirow{2}{*}{Radiative decay channel}&\multicolumn{3}{c}{$\Gamma$ (keV)}\\\cline{3-5}
		\multirow{12}{*}{$cc\bar{n}\bar{s}$}&
		&$\langle\mu\rangle^2$&$I_3=\frac12$&$I_3=-\frac12$\\\cline{1-5}
		&$T_{cc}^{(\frac12,2)}(4218)\rightarrow T_{cc}^{(\frac12,1)}(4182)\gamma$&\multirow{3}{*}{\begin{tabular}{c}$\frac56[\beta_1(2\mu_c-\mu_{\bar{n}}-\mu_{\bar{s}})$\\$+\sqrt{2}\beta_2(\mu_{\bar{n}}-\mu_{\bar{s}})]^2$\end{tabular}}& {$2.1\times 10^{-3}$}&0.19\\
		&$T_{cc}^{(\frac12,2)}(4218)\rightarrow T_{cc}^{(\frac12,1)}(4147)\gamma$& &0.41& 1.81\\
		&$T_{cc}^{(\frac12,2)}(4218)\rightarrow T_{cc}^{(\frac12,1)}(4009)\gamma$& &1.33& 111.16\\\cline{3-5}
		&$T_{cc}^{(\frac12,1)}(4182)\rightarrow T_{cc}^{(\frac12,1)}(4147)\gamma$&\multirow{3}{*}{\begin{tabular}{c}$[\alpha_1\beta_1(2\mu_c+\mu_{\bar{n}}+\mu_{\bar{s}})+$\\$4\alpha_2\beta_2\mu_c+2\alpha_3\beta_3(\mu_{\bar{n}}+\mu_{\bar{s}})+$\\$\sqrt{2}(\alpha_1\beta_2+\alpha_2\beta_1)(\mu_{\bar{n}}-\mu_{\bar{s}})]^2/2$\end{tabular}}& {$2.41\times 10^{-3}$}&0.20\\
		&$T_{cc}^{(\frac12,1)}(4182)\rightarrow T_{cc}^{(\frac12,1)}(4009)\gamma$& &4.27& 17.20\\
		&$T_{cc}^{(\frac12,1)}(4147)\rightarrow T_{cc}^{(\frac12,1)}(4009)\gamma$& & {0.43}& 33.51\\\cline{3-5}
		&$T_{cc}^{(\frac12,1)}(4182)\rightarrow T_{cc}^{(\frac12,0)}(4018)\gamma$&\multirow{2}{*}{\begin{tabular}{c}$\frac19[\sqrt{6}\alpha_1\beta_1(2\mu_c-\mu_{\bar{n}}-\mu_{\bar{s}})-$\\$(\sqrt{3}\alpha_2\beta_1-3\alpha_3\beta_2)(\mu_{\bar{n}}-\mu_{\bar{s}})]^2$\end{tabular}}&0.05&4.79\\
		&$T_{cc}^{(\frac12,1)}(4147)\rightarrow T_{cc}^{(\frac12,0)}(4018)\gamma$&&2.08&8.83\\\cline{3-5}
		&$T_{cc}^{(\frac12,0)}(4295)\rightarrow T_{cc}^{(\frac12,1)}(4182)\gamma$&\multirow{4}{*}{\begin{tabular}{c}$\frac19[\sqrt{6}\alpha_1\beta_1(2\mu_c-\mu_{\bar{n}}-\mu_{\bar{s}})-$\\$(\sqrt{3}\alpha_1\beta_2-3\alpha_2\beta_3)(\mu_{\bar{n}}-\mu_{\bar{s}})]^2$\end{tabular}} &0.73&57.17\\
		&$T_{cc}^{(\frac12,0)}(4295)\rightarrow T_{cc}^{(\frac12,1)}(4147)\gamma$& &4.81& 19.38\\
		&$T_{cc}^{(\frac12,0)}(4295)\rightarrow T_{cc}^{(\frac12,1)}(4009)\gamma$& &0.18& 16.46\\
		&$T_{cc}^{(\frac12,0)}(4019)\rightarrow T_{cc}^{(\frac12,1)}(4009)\gamma$& &$2.90\times 10^{-4}$ &0.02\\
		\hline
		\multirow{12}{*}{$bb\bar{n}\bar{s}$}
		&$T_{bb}^{(\frac12,2)}(10870)\rightarrow T_{bb}^{(\frac12,1)}(10847)\gamma$&\multirow{3}{*}{\begin{tabular}{c}$\frac56[\beta_1(2\mu_b-\mu_{\bar{n}}-\mu_{\bar{s}})$\\$+\sqrt{2}\beta_2(\mu_{\bar{n}}-\mu_{\bar{s}})]^2$\end{tabular}}&0.06&0.02\\
		&$T_{bb}^{(\frac12,2)}(10870)\rightarrow T_{bb}^{(\frac12,1)}(10819)\gamma$& &0.01& 0.18\\
		&$T_{bb}^{(\frac12,2)}(10870)\rightarrow T_{bb}^{(\frac12,1)}(10722)\gamma$& &0.59& 52.12\\\cline{3-5}
		&$T_{bb}^{(\frac12,1)}(10847)\rightarrow T_{bb}^{(\frac12,1)}(10819)\gamma$&\multirow{3}{*}{\begin{tabular}{c}$\frac12[\alpha_1\beta_1(2\mu_b+\mu_{\bar{n}}+\mu_{\bar{s}})+$\\$4\alpha_2\beta_2\mu_b+2\alpha_3\beta_3(\mu_{\bar{n}}+\mu_{\bar{s}})+$\\$\sqrt{2}(\alpha_1\beta_2+\alpha_2\beta_1)(\mu_{\bar{n}}-\mu_{\bar{s}})]^2$\end{tabular}}& {$1.02\times 10^{-3}$}&0.03\\
		&$T_{bb}^{(\frac12,1)}(10847)\rightarrow T_{bb}^{(\frac12,1)}(10722)\gamma$& &0.32& 30.94\\
		&$T_{bb}^{(\frac12,1)}(10819)\rightarrow T_{bb}^{(\frac12,1)}(10722)\gamma$& &1.09& 0.59\\\cline{3-5}
		&$T_{bb}^{(\frac12,1)}(10847)\rightarrow T_{bb}^{(\frac12,0)}(10808)\gamma$&\multirow{2}{*}{\begin{tabular}{c}$\frac19[\sqrt{6}\alpha_1\beta_1(2\mu_b-\mu_{\bar{n}}-\mu_{\bar{s}})-$\\$(\sqrt{3}\alpha_2\beta_1-3\alpha_3\beta_2)(\mu_{\bar{n}}-\mu_{\bar{s}})]^2$\end{tabular}}&0.24&0.07\\
		&$T_{bb}^{(\frac12,1)}(10819)\rightarrow T_{bb}^{(\frac12,0)}(10808)\gamma$&&$1.10\times10^{-4}$&{$3.98\times 10^{-3}$}\\\cline{3-5}	
		&$T_{bb}^{(\frac12,0)}(10904)\rightarrow T_{bb}^{(\frac12,1)}(10847)\gamma$&\multirow{4}{*}{\begin{tabular}{c}$\frac19[\sqrt{6}\alpha_1\beta_1(2\mu_c-\mu_{\bar{n}}-\mu_{\bar{s}})-$\\$(\sqrt{3}\alpha_1\beta_2-3\alpha_2\beta_3)(\mu_{\bar{n}}-\mu_{\bar{s}})]^2$\end{tabular}}&0.95&0.48\\
		&$T_{bb}^{(\frac12,0)}(10904)\rightarrow T_{bb}^{(\frac12,1)}(10819)\gamma$&&0.23& 25.47\\
		&$T_{bb}^{(\frac12,0)}(10904)\rightarrow T_{bb}^{(\frac12,1)}(10722)\gamma$&&0.03& 1.55\\
		&$T_{bb}^{(\frac12,0)}(10808)\rightarrow T_{bb}^{(\frac12,1)}(10722)\gamma$&&0.17& 12.60\\
		\hline\hline
	\end{tabular}	
\end{table}

We present the results for radiative transitions of the $QQ\bar{n}\bar{s}$ states in Table \ref{tab:QQns-radiative-width}, which depend on the third component of isospin $I_3$.
From the third column, it is evident that the amplitudes of all the decay channels encompass the spin-color structure weights $\alpha_i$ or $\beta_i$, which indicates that the inner structure information of the $QQ\bar{n}\bar{s}$ states can be extracted from their electromagnetic radiative processes.
Especially, the decays between the three $J=1$ states are always influenced by the coupling of different spin configurations, as reflected in the cross terms such as $\sqrt{2}(\alpha_1\beta_2+\alpha_2\beta_1)(\mu_{\bar{n}}-\mu_{\bar{s}})$.
Such a term suggests the spin-flip transition between the $S_{\bar{n}\bar{s}}=1$ and $S_{\bar{n}\bar{s}}=0$ states.

In the $cc\bar{n}\bar{s}$ system, the most prominent radiative decay channel is $T_{cc}^{(\frac12,2)}(4218)\rightarrow T_{cc}^{(\frac12,1)}(4009)\gamma$, with a width of $111.16$ keV, while the weakest channel is $T_{cc}^{(\frac12,0)}(4019)\rightarrow T_{cc}^{(\frac12,1)}(4009)\gamma$, with a width of $2.90\times 10^{-4}$ keV.
In the $bb\bar{n}\bar{s}$ system, the broadest radiative decay channel is $T_{bb}^{(\frac12,2)}(10870)\rightarrow T_{bb}^{(\frac12,1)}(10722)\gamma$, with a width of $52.12$ keV, whereas the narrowest channel is $T_{bb}^{(\frac12,1)}(10819)\rightarrow T_{bb}^{(\frac12,0)}(10808)\gamma$, with a width of $1.10\times10^{-4}$ keV.
A general feature is that, in the $cc\bar{n}\bar{s}$ and $bb\bar{n}\bar{s}$ systems, the widths for the radiative transitions in the $I_3=\frac12$ case are smaller than those in the $I_3=-\frac12$ case.
However, this trend is reversed in four specific radiative decay channels of the $bb\bar{n}\bar{s}$ system: $T^{(\frac12,1)}(10870)\rightarrow T_{bb}^{(\frac12,1)}(10847)\gamma$, $T_{bb}^{(\frac12,1)}(10819)\rightarrow T_{bb}^{(\frac12,1)}(10722)\gamma$, $T_{bb}^{(\frac12,1)}(10847)\rightarrow T^{(\frac12,1)}(10808)\gamma$, and $T^{(\frac12,1)}(10904)\rightarrow T_{bb}^{(\frac12,1)}(10847)\gamma$.

\begin{table}[htbp]
	\caption{Radiative transitions of $bc\bar{n}\bar{s}$ states in units of keV. A tetraquark is denoted as $T^{(I,J)}(Mass)$. In the second column, $\alpha_i$ and $\beta_i$ represent the weight coefficients of the spin-color configurations in the initial and final states, respectively.}\label{tab:bcns-radiative-width}
	\footnotesize
	\begin{tabular}{c|c|cccccc}\hline\hline
		\multirow{2}{*}{Radiative decay channel}&\multicolumn{3}{c}{$\Gamma$ (keV)}\\\cline{2-4}
		&$\langle\mu\rangle^2$&$I_3=\frac12$&$I_3=-\frac12$\\\hline
		$T_{bc}^{(\frac12,2)}(7542)\rightarrow T_{bc}^{(\frac12,1)}(7511)\gamma$&\multirow{10}{*}{\begin{tabular}{c}$\frac56[(\alpha_1\beta_1+\alpha_2\beta_2)(\mu_b+\mu_c-\mu_{\bar{n}}-\mu_{\bar{s}})+$\\$\sqrt{2}\alpha_1(\beta_5(\mu_b-\mu_c)+\beta_3(\mu_{\bar{n}}-\mu_{\bar{s}}))+$\\$\sqrt{2}\alpha_2(\beta_6(\mu_b-\mu_c)+\beta_4(\mu_{\bar{n}}-\mu_{\bar{s}}))]^2$\end{tabular}} &0.08&0.02\\
		$T_{bc}^{(\frac12,2)}(7542)\rightarrow T_{bc}^{(\frac12,1)}(7502)\gamma$& &0.01& 0.14\\
		$T_{bc}^{(\frac12,2)}(7542)\rightarrow T_{bc}^{(\frac12,1)}(7437)\gamma$& &0.12& 0.85\\
		$T_{bc}^{(\frac12,2)}(7542)\rightarrow T_{bc}^{(\frac12,1)}(7432)\gamma$& &0.14& 11.02\\
		$T_{bc}^{(\frac12,2)}(7542)\rightarrow T_{bc}^{(\frac12,1)}(7330)\gamma$& &0.99& 79.52\\
		$T_{bc}^{(\frac12,2)}(7521)\rightarrow T_{bc}^{(\frac12,1)}(7511)\gamma$& &$4.2\times10^{-5}$&  {$1.1\times10^{-4}$}\\
		$T_{bc}^{(\frac12,2)}(7521)\rightarrow T_{bc}^{(\frac12,1)}(7502)\gamma$& &0.01&  {$5.8\times10^{-4}$}\\
		$T_{bc}^{(\frac12,2)}(7521)\rightarrow T_{bc}^{(\frac12,1)}(7437)\gamma$& &0.15& 3.96\\
		$T_{bc}^{(\frac12,2)}(7521)\rightarrow T_{bc}^{(\frac12,1)}(7432)\gamma$& &0.91& 0.49\\
		$T_{bc}^{(\frac12,2)}(7521)\rightarrow T_{bc}^{(\frac12,1)}(7330)\gamma$& &2.61& 10.33\\\cline{2-4}
		$T_{bc}^{(\frac12,1)}(7569)\rightarrow T_{bc}^{(\frac12,2)}(7542)\gamma$&\begin{tabular}{c}$\frac56[(\alpha_1\beta_1+\alpha_2\beta_2)(\mu_b+\mu_c-\mu_{\bar{n}}-\mu_{\bar{s}})+$\\$\sqrt{2}\beta_1(\alpha_5(\mu_b-\mu_c)+\alpha_3(\mu_{\bar{n}}-\mu_{\bar{s}}))+$\\$\sqrt{2}\beta_2(\alpha_6(\mu_b-\mu_c)+\alpha_4(\mu_{\bar{n}}-\mu_{\bar{s}}))]^2$\end{tabular}&$2.8\times10^{-4}$&0.02\\
		$T_{bc}^{(\frac12,1)}(7569)\rightarrow T_{bc}^{(\frac12,2)}(7521)\gamma$& &0.03& 2.22\\\cline{2-4}
		$T_{bc}^{(\frac12,1)}(7569)\rightarrow T_{bc}^{(\frac12,1)}(7511)\gamma$&\multirow{15}{*}{\begin{tabular}{c}$\frac12[(\alpha^2_1+\alpha^2_2)(\mu_b+\mu_c+\mu_{\bar{n}}+\mu_{\bar{s}})+$\\$2(\alpha^2_3+\alpha^2_4)(\mu_b+\mu_c)+$\\$2(\alpha^2_5+\alpha^2_6)(\mu_{\bar{n}}+\mu_{\bar{s}})+$\\$2\sqrt{2}(\alpha_1\alpha_5+\alpha_2\alpha_6)(\mu_c-\mu_b)+$\\$2\sqrt{2}(\alpha_1\alpha_3+\alpha_2\alpha_4)(\mu_{\bar{n}}-\mu_{\bar{s}})]^2$\end{tabular}}&0.11&0.06\\
		$T_{bc}^{(\frac12,1)}(7569)\rightarrow T_{bc}^{(\frac12,1)}(7502)\gamma$&& {$1.0\times 10^{-3}$}&0.06\\
		$T_{bc}^{(\frac12,1)}(7569)\rightarrow T_{bc}^{(\frac12,1)}(7437)\gamma$&&2.15&12.25\\
		$T_{bc}^{(\frac12,1)}(7569)\rightarrow T_{bc}^{(\frac12,1)}(7432)\gamma$&&0.07&22.89\\
		$T_{bc}^{(\frac12,1)}(7569)\rightarrow T_{bc}^{(\frac12,1)}(7330)\gamma$&&0.21&18.05\\
		$T_{bc}^{(\frac12,1)}(7511)\rightarrow T_{bc}^{(\frac12,1)}(7502)\gamma$&&$1.90\times10^{-5}$& {$8.7\times 10^{-4}$}\\
		$T_{bc}^{(\frac12,1)}(7511)\rightarrow T_{bc}^{(\frac12,1)}(7437)\gamma$&&0.14&0.02\\
		$T_{bc}^{(\frac12,1)}(7511)\rightarrow T_{bc}^{(\frac12,1)}(7432)\gamma$&&0.05&2.90\\
		$T_{bc}^{(\frac12,1)}(7511)\rightarrow T_{bc}^{(\frac12,1)}(7330)\gamma$&&0.15&22.83\\
		$T_{bc}^{(\frac12,1)}(7502)\rightarrow T_{bc}^{(\frac12,1)}(7437)\gamma$&& {$7.5\times 10^{-4}$}&0.48\\
		$T_{bc}^{(\frac12,1)}(7502)\rightarrow T_{bc}^{(\frac12,1)}(7432)\gamma$&&0.27&0.07\\
		$T_{bc}^{(\frac12,1)}(7502)\rightarrow T_{bc}^{(\frac12,1)}(7330)\gamma$&&2.24&12.06\\
		$T_{bc}^{(\frac12,1)}(7437)\rightarrow T_{bc}^{(\frac12,1)}(7432)\gamma$&&$1.2\times10^{-6}$&$4.3\times10^{-5}$\\
		$T_{bc}^{(\frac12,1)}(7437)\rightarrow T_{bc}^{(\frac12,1)}(7330)\gamma$&&0.06&8.76\\
		$T_{bc}^{(\frac12,1)}(7432)\rightarrow T_{bc}^{(\frac12,1)}(7330)\gamma$&&0.30&2.49\\\cline{2-4}
		$T_{bc}^{(\frac12,1)}(7569)\rightarrow T_{bc}^{(\frac12,0)}(7444)\gamma$&\multirow{14}{*}{\begin{tabular}{c}$\frac19[\sqrt{6}(\alpha_1\beta_1+\alpha_2\beta_2)(\mu_b+\mu_c-\mu_{\bar{n}}-\mu_{\bar{s}})+$\\$(3(\alpha_3\beta_3+\alpha_4\beta_4)-\sqrt{3}(\alpha_5\beta_1+\alpha_6\beta_2))(\mu_b-\mu_c)+$\\$(3(\alpha_5\beta_3+\alpha_6\beta_4)-\sqrt{3}(\alpha_3\beta_1+\alpha_4\beta_2))(\mu_{\bar{n}}-\mu_{\bar{s}})]^2$\end{tabular}}&0.18&15.75\\
		$T_{bc}^{(\frac12,1)}(7569)\rightarrow T_{bc}^{(\frac12,0)}(7411)\gamma$&&0.27&1.60\\
		$T_{bc}^{(\frac12,1)}(7569)\rightarrow T_{bc}^{(\frac12,0)}(7269)\gamma$&& {$2.1\times 10^{-3}$}&0.35\\
		$T_{bc}^{(\frac12,1)}(7511)\rightarrow T_{bc}^{(\frac12,0)}(7444)\gamma$&&0.01&1.04\\
		$T_{bc}^{(\frac12,1)}(7511)\rightarrow T_{bc}^{(\frac12,0)}(7411)\gamma$&&0.87&1.65\\
		$T_{bc}^{(\frac12,1)}(7511)\rightarrow T_{bc}^{(\frac12,0)}(7269)\gamma$&&0.36&39.47\\
		$T_{bc}^{(\frac12,1)}(7502)\rightarrow T_{bc}^{(\frac12,0)}(7444)\gamma$&& {$2.9\times 10^{-3}$}&0.06\\
		$T_{bc}^{(\frac12,1)}(7502)\rightarrow T_{bc}^{(\frac12,0)}(7411)\gamma$&&0.06&2.21\\
		$T_{bc}^{(\frac12,1)}(7502)\rightarrow T_{bc}^{(\frac12,0)}(7269)\gamma$&&0.81&6.88\\
		$T_{bc}^{(\frac12,1)}(7437)\rightarrow T_{bc}^{(\frac12,0)}(7411)\gamma$&&0.03& {$7.9\times 10^{-3}$}\\
		$T_{bc}^{(\frac12,1)}(7437)\rightarrow T_{bc}^{(\frac12,0)}(7269)\gamma$&&0.08&27.49\\
		$T_{bc}^{(\frac12,1)}(7432)\rightarrow T_{bc}^{(\frac12,0)}(7411)\gamma$&&$6.1\times10^{-5}$&0.01\\
		$T_{bc}^{(\frac12,1)}(7432)\rightarrow T_{bc}^{(\frac12,0)}(7269)\gamma$&&2.70&16.98\\
		$T_{bc}^{(\frac12,1)}(7330)\rightarrow T_{bc}^{(\frac12,0)}(7269)\gamma$&&0.36&0.04\\\cline{2-4}
		$T_{bc}^{(\frac12,0)}(7596)\rightarrow T_{bc}^{(\frac12,1)}(7569)\gamma$&\multirow{10}{*}{\begin{tabular}{c}$\frac19[\sqrt{6}(\alpha_1\beta_1+\alpha_2\beta_2)(\mu_b+\mu_c-\mu_{\bar{n}}-\mu_{\bar{s}})+$\\$(3(\alpha_3\beta_3+\alpha_4\beta_4)-\sqrt{3}(\alpha_1\beta_5+\alpha_2\beta_6))(\mu_b-\mu_c)+$\\$(3(\alpha_3\beta_5+\alpha_4\beta_6)-\sqrt{3}(\alpha_1\beta_3+\alpha_2\beta_4))(\mu_{\bar{n}}-\mu_{\bar{s}})]^2$\end{tabular}}&0.05& {$3.5\times 10^{-3}$}\\
		$T_{bc}^{(\frac12,0)}(7596)\rightarrow T_{bc}^{(\frac12,1)}(7511)\gamma$&&1.11&3.63\\
		$T_{bc}^{(\frac12,0)}(7596)\rightarrow T_{bc}^{(\frac12,1)}(7502)\gamma$&&0.21&29.32\\
		$T_{bc}^{(\frac12,0)}(7596)\rightarrow T_{bc}^{(\frac12,1)}(7437)\gamma$&&1.69&10.63\\
		$T_{bc}^{(\frac12,0)}(7596)\rightarrow T_{bc}^{(\frac12,1)}(7432)\gamma$&&0.02&6.83\\
		$T_{bc}^{(\frac12,0)}(7596)\rightarrow T_{bc}^{(\frac12,1)}(7330)\gamma$&&0.13&11.91\\
		$T_{bc}^{(\frac12,0)}(7444)\rightarrow T_{bc}^{(\frac12,1)}(7437)\gamma$&&$1.30\times 10^{-4}$& {$1.5\times 10^{-3}$}\\
		$T_{bc}^{(\frac12,0)}(7444)\rightarrow T_{bc}^{(\frac12,1)}(7432)\gamma$&& {$7.8\times 10^{-3}$}&$8.10\times 10^{-4}$\\
		$T_{bc}^{(\frac12,0)}(7444)\rightarrow T_{bc}^{(\frac12,1)}(7330)\gamma$&&0.87&4.79\\
		$T_{bc}^{(\frac12,0)}(7411)\rightarrow T_{bc}^{(\frac12,1)}(7330)\gamma$&&0.12&8.98\\
		\hline\hline
	\end{tabular}	
\end{table}

\subsubsection{The $bc\bar{n}\bar{s}$ system}

\begin{table}[!h]
	\centering
	\caption{Radiative transitions of $bc\bar{n}\bar{s}$ states with $\Gamma(I_3=\frac{1}{2}) > \Gamma(I_3=-\frac{1}{2})$ and the corresponding width ratios.}\label{tab:ratio-bcns}
	\begin{tabular}{c|c|c|c}
		\hline\hline
		Radiative decay channel & $\Gamma(I_3=\frac{1}{2})$ (keV) & $\Gamma(I_3=-\frac{1}{2})$ (keV) & Ratio \\
		\hline
		$T_{bc}^{(\frac{1}{2},2)}(7542) \to T_{bc}^{(\frac{1}{2},1)}(7511)\gamma$ & 0.08 & 0.02 & 4.15 \\
		$T_{bc}^{(\frac{1}{2},2)}(7521) \to T_{bc}^{(\frac{1}{2},1)}(7502)\gamma$ & 0.01 & $5.8\times10^{-4}$ & 17.24 \\
		$T_{bc}^{(\frac{1}{2},2)}(7521) \to T_{bc}^{(\frac{1}{2},1)}(7432)\gamma$ & 0.91 & 0.49 & 1.87 \\
		$T_{bc}^{(\frac{1}{2},1)}(7569) \to T_{bc}^{(\frac{1}{2},1)}(7511)\gamma$ & 0.11 & 0.06 & 1.71 \\
		$T_{bc}^{(\frac{1}{2},1)}(7511) \to T_{bc}^{(\frac{1}{2},1)}(7437)\gamma$ & 0.14 & 0.02 & 8.35 \\
		$T_{bc}^{(\frac{1}{2},1)}(7502) \to T_{bc}^{(\frac{1}{2},1)}(7432)\gamma$ & 0.27 & 0.07 & 3.97 \\
		$T_{bc}^{(\frac{1}{2},1)}(7437) \to T_{bc}^{(\frac{1}{2},0)}(7411)\gamma$ & 0.03 & $7.9\times10^{-3}$ & 3.80 \\
		$T_{bc}^{(\frac{1}{2},1)}(7330) \to T_{bc}^{(\frac{1}{2},0)}(7269)\gamma$ & 0.36 & 0.04 & 8.83 \\
		$T_{bc}^{(\frac{1}{2},0)}(7596) \to T_{bc}^{(\frac{1}{2},1)}(7569)\gamma$ & 0.05 & $3.5\times10^{-3}$ & 14.29 \\
		$T_{bc}^{(\frac{1}{2},0)}(7444) \to T_{bc}^{(\frac{1}{2},1)}(7432)\gamma$ & $7.8\times 10^{-3}$ & $8.10\times 10^{-4}$& 9.63 \\
		\hline\hline
	\end{tabular}
\end{table}

In the $bc\bar{n}\bar{s}$ system, numerous radiative decay channels exist, as detailed in Table \ref{tab:bcns-radiative-width}.
We observe that the radiative decay widths for the $I_3=-\frac12$ case are generally larger than those for the $I_3=\frac12$ case.
However, the complex interactions within the $bc\bar{n}\bar{s}$ states lead to an unexpected outcome: ten radiative decay channels have larger widths for the $I_3=\frac12$ case than for the $I_3=-\frac12$ case. We collect results of such channels in table \ref{tab:ratio-bcns}.
Those ten channels can serve as indicators of the compact $bc\bar{n}\bar{s}$ tetraquark states.

The range of radiative decay widths in the $bc\bar{n}\bar{s}$ system is relatively wide.
The most prominent radiative decay channel in the $bc\bar{n}\bar{s}$ system is $T_{bc}^{(\frac12,2)}(7542)\rightarrow T_{bc}^{(\frac12,1)}(7330)\gamma$, with a width of $79.52$ keV.
This decay channel may be more significant in experimental searches for $bc\bar{n}\bar{s}$ tetraquark states.
Conversely, the weakest channel is $T_{bc}^{(\frac12,1)}(7437)\rightarrow T_{bc}^{(\frac12,1)}(7432)\gamma$, with a width of $1.20\times10^{-6}$ keV.
For a given radiative decay channel, the widths of $I_3=-\frac12$ and $I_3=\frac12$ show a significant difference, e.g., in the channel $T_{bc}^{(\frac12,2)}(7542)\rightarrow T_{bc}^{(\frac12,1)}(7330)\gamma$, the width of $I_3=-\frac12$ is about $80$ times that of $I_3=\frac12$.

\begin{table}[htbp]
	\caption{Radiative transitions of $bc\bar{n}\bar{n}^\prime$ states in units of keV. A tetraquark is denoted as $T^{(I,J)}(Mass)$. In the second column, $\alpha_i$ and $\beta_i$ represent the weight coefficients of the spin-color configurations in the initial and final states, respectively.}\label{tab:bcnn-radiative-width}
	\footnotesize
	\begin{tabular}{c|c|cccccc}\hline\hline
		\multirow{2}{*}{Radiative decay channel}&\multicolumn{4}{c}{$\Gamma$ (keV)}\\\cline{2-5}
		&$\langle\mu\rangle^2$&$I_3=1$&$I_3=0$&$I_3=-1$\\\cline{1-5}
		$T_{bc}^{(1,2)}(7468)\rightarrow T_{bc}^{(1,1)}(7438)\gamma$&\multirow{2}{*}{\begin{tabular}{c}$\frac56[\beta_1(\mu_b+\mu_c-\mu_{\bar{n}}-\mu_{\bar{n}^\prime})$\\$+\sqrt{2}\beta_3(\mu_b-\mu_c))]^2$\end{tabular}}& {0.10}&0.01& {0.29}\\
		$T_{bc}^{(1,2)}(7468)\rightarrow T_{bc}^{(1,1)}(7366)\gamma$& & {0.35}&1.56&9.58\\\cline{2-5}
		$T_{bc}^{(1,2)}(7468)\rightarrow T_{bc}^{(0,1)}(7401)\gamma$&\multirow{3}{*}{$\frac53\beta_2^2(\mu_{\bar{d}}-\mu_{\bar{u}})^2$}&&0.32&\\
		$T_{bc}^{(1,2)}(7468)\rightarrow T_{bc}^{(0,1)}(7320)\gamma$&&&20.00&\\
		$T_{bc}^{(1,2)}(7468)\rightarrow T_{bc}^{(0,1)}(7208)\gamma$&&&244.15&\\\cline{2-5}
		$T_{bc}^{(0,2)}(7421)\rightarrow T_{bc}^{(1,1)}(7366)\gamma$&$\frac53\beta_2^2(\mu_{\bar{d}}-\mu_{\bar{u}})^2$&&1.06&\\\cline{2-5}
		$T_{bc}^{(0,2)}(7421)\rightarrow T_{bc}^{(0,1)}(7401)\gamma$&\multirow{3}{*}{\begin{tabular}{c}$\frac56[\beta_1(\mu_b+\mu_c-\mu_{\bar{d}}-\mu_{\bar{u}})$\\$+\sqrt{2}\beta_3(\mu_b-\mu_c))]^2$\end{tabular}}&& {$3.37\times 10^{-4}$}&\\
		$T_{bc}^{(0,2)}(7421)\rightarrow T_{bc}^{(0,1)}(7320)\gamma$&&&  {1.63}&\\
		$T_{bc}^{(0,2)}(7421)\rightarrow T_{bc}^{(0,1)}(7208)\gamma$&&&  {7.43}&\\\cline{2-5}
		$T_{bc}^{(1,1)}(7500)\rightarrow T_{bc}^{(1,2)}(7468)\gamma$&\begin{tabular}{c}$\frac56[\alpha_1(\mu_b+\mu_c-\mu_{\bar{n}}-\mu_{\bar{n}^\prime})$\\$+\sqrt{2}\alpha_3(\mu_b-\mu_c))]^2$\end{tabular}& {$2.48\times 10^{-3}$}&0.03&0.15\\\cline{2-5}
		$T_{bc}^{(1,1)}(7500)\rightarrow T_{bc}^{(0,2)}(7421)\gamma$&\multirow{2}{*}{$\frac53\alpha_2^2(\mu_{\bar{d}}-\mu_{\bar{u}})^2$}&&12.59&\\
		$T_{bc}^{(1,1)}(7438)\rightarrow T_{bc}^{(0,2)}(7421)\gamma$&&&  {$8.48\times 10^{-4}$}&\\\cline{2-5}
		$T_{bc}^{(1,1)}(7500)\rightarrow T_{bc}^{(1,1)}(7438)\gamma$&\multirow{3}{*}{\begin{tabular}{c}$\frac12[\alpha_1\beta_1(\mu_b+\mu_c+\mu_{\bar{n}}+\mu_{\bar{n}^\prime})+$\\$2\alpha_2\beta_2(\mu_b+\mu_c)+2\alpha_3\beta_3(\mu_{\bar{n}}+\mu_{\bar{n}^\prime})+$\\$\sqrt{2}(\alpha_3\beta_1+\alpha_1\beta_3)(\mu_c-\mu_b)]^2$\end{tabular}}&0.15&0.03&0.51\\
		$T_{bc}^{(1,1)}(7500)\rightarrow T_{bc}^{(1,1)}(7366)\gamma$&&3.02&4.36&34.99\\
		$T_{bc}^{(1,1)}(7438)\rightarrow T_{bc}^{(1,1)}(7366)\gamma$&&0.24&0.05&0.88\\\cline{2-5}
		$T_{bc}^{(1,1)}(7500)\rightarrow T_{bc}^{(0,1)}(7401)\gamma$&\multirow{8}{*}{$(\alpha_1\beta_2+\alpha_2\beta_1)^2(\mu_{\bar{d}}-\mu_{\bar{u}})^2$}&&0.88&\\
		$T_{bc}^{(1,1)}(7500)\rightarrow T_{bc}^{(0,1)}(7320)\gamma$&&&89.13&\\
		$T_{bc}^{(1,1)}(7500)\rightarrow T_{bc}^{(0,1)}(7208)\gamma$&&&11.97&\\
		$T_{bc}^{(1,1)}(7438)\rightarrow T_{bc}^{(0,1)}(7401)\gamma$&&&0.03&\\
		$T_{bc}^{(1,1)}(7438)\rightarrow T_{bc}^{(0,1)}(7320)\gamma$&&&8.15&\\
		$T_{bc}^{(1,1)}(7438)\rightarrow T_{bc}^{(0,1)}(7208)\gamma$&&& {95.84}&\\
		$T_{bc}^{(1,1)}(7366)\rightarrow T_{bc}^{(0,1)}(7320)\gamma$&&&0.05&\\
		$T_{bc}^{(1,1)}(7366)\rightarrow T_{bc}^{(0,1)}(7208)\gamma$&&&39.93&\\\cline{2-5}
		$T_{bc}^{(1,1)}(7500)\rightarrow T_{bc}^{(1,0)}(7342)\gamma$&\multirow{3}{*}{\begin{tabular}{c}$\frac19[(\sqrt{3}\alpha_3\beta_1-3\alpha_2\beta_2)(\mu_b-\mu_c)-$\\$\sqrt{6}\alpha_1\beta_1(\mu_b+\mu_c-\mu_{\bar{n}}-\mu_{\bar{n}})]^2$\end{tabular}}&0.49&1.16&8.16\\
		$T_{bc}^{(1,1)}(7438)\rightarrow T_{bc}^{(1,0)}(7342)\gamma$& &1.29&1.45&12.53\\
		$T_{bc}^{(1,1)}(7366)\rightarrow T_{bc}^{(1,0)}(7342)\gamma$& &0.03& {$1.09\times 10^{-3}$}& 0.05\\\cline{2-5}
		$T_{bc}^{(1,1)}(7500)\rightarrow T^{(0,0)}(7322)\gamma$&\multirow{6}{*}{\begin{tabular}{c}$(\sqrt{3}\alpha_2\beta_1-3\alpha_3\beta_2)^2$\\$(\mu_{\bar{d}}-\mu_{\bar{u}})^2/9$\end{tabular}}&&53.41&\\
		$T_{bc}^{(1,1)}(7500)\rightarrow T_{bc}^{(0,0)}(7151)\gamma$&&& 0.60&\\
		$T_{bc}^{(1,1)}(7438)\rightarrow T^{(0,0)}(7322)\gamma$&&& 4.65&\\
		$T_{bc}^{(1,1)}(7438)\rightarrow T_{bc}^{(0,0)}(7151)\gamma$&&& 114.53&\\
		$T_{bc}^{(1,1)}(7366)\rightarrow T^{(0,0)}(7322)\gamma$&&& 0.07&\\
		$T_{bc}^{(1,1)}(7366)\rightarrow T_{bc}^{(0,0)}(7151)\gamma$&&& 99.36&\\\cline{2-5}		
		$T_{bc}^{(0,1)}(7401)\rightarrow T_{bc}^{(1,1)}(7366)\gamma$&$(\alpha_1\beta_2+\alpha_2\beta_1)^2(\mu_{\bar{d}}-\mu_{\bar{u}})^2$&&0.09&\\\cline{2-5}
		$T_{bc}^{(0,1)}(7401)\rightarrow T_{bc}^{(0,1)}(7320)\gamma$&\multirow{3}{*}{\begin{tabular}{c}$\frac12[\alpha_1\beta_1(\mu_b+\mu_c+\mu_{\bar{d}}+\mu_{\bar{u}})+$\\$2\alpha_3\beta_3(\mu_b+\nu_c)+$\\$\sqrt{2}(\alpha_3\beta_1+\alpha_1\beta_3)(\mu_c-\mu_b)]^2$ \end{tabular}} &&0.16&\\
		$T_{bc}^{(0,1)}(7401)\rightarrow T_{bc}^{(0,1)}(7208)\gamma$&&& 6.04&\\
		$T_{bc}^{(0,1)}(7320)\rightarrow T_{bc}^{(0,1)}(7208)\gamma$&&& 1.26&\\\cline{2-5}
		$T_{bc}^{(0,1)}(7401)\rightarrow T_{bc}^{(1,0)}(7342)\gamma$&$(\sqrt{3}\alpha_2\beta_1-3\alpha_3\beta_2)^2(\mu_{\bar{d}}-\mu_{\bar{u}})^2/9$&& {0.70}&\\\cline{2-5}
		$T_{bc}^{(0,1)}(7401)\rightarrow T^{(0,0)}(7322)\gamma$&\multirow{4}{*}{\begin{tabular}{c}$\frac19[\sqrt{6}\alpha_1\beta_1(\mu_{\bar{d}}+\mu_{\bar{u}}-\mu_b-\mu_c)+$\\$(\sqrt{3}\alpha_3\beta_1-3\alpha_2\beta_2)(\mu_b-\mu_c)]^2$\end{tabular}} &&0.29&\\
		$T_{bc}^{(0,1)}(7401)\rightarrow T_{bc}^{(0,0)}(7151)\gamma$&&& 3.42&\\
		$T_{bc}^{(0,1)}(7320)\rightarrow T_{bc}^{(0,0)}(7151)\gamma$&&& 5.98&\\
		$T_{bc}^{(0,1)}(7208)\rightarrow T_{bc}^{(0,0)}(7151)\gamma$&&& {$4.16\times 10^{-4}$} &&\\\cline{2-5}
		$T_{bc}^{(1,0)}(7528)\rightarrow T_{bc}^{(1,1)}(7500)\gamma$&\multirow{3}{*}{\begin{tabular}{c}$\frac19[(\sqrt{3}\alpha_1\beta_3-3\alpha_2\beta_2)(\mu_b-\mu_c)-$\\$\sqrt{6}\alpha_1\beta_1(\mu_b+\mu_c-\mu_{\bar{n}}-\mu_{\bar{n}})]^2$\end{tabular}}&0.06& {$8.67\times 10^{-3}$}& {$3.96\times 10^{-3}$}\\
		$T_{bc}^{(1,0)}(7528)\rightarrow T_{bc}^{(1,1)}(7438)\gamma$& &1.63&1.38&13.15\\
		$T_{bc}^{(1,0)}(7528)\rightarrow T_{bc}^{(1,1)}(7366)\gamma$& &2.92&5.40&40.42\\\cline{2-5}
		$T_{bc}^{(1,0)}(7528)\rightarrow T_{bc}^{(0,1)}(7401)\gamma$&\multirow{5}{*}{\begin{tabular}{c}$(\sqrt{3}\alpha_1\beta_2-3\alpha_2\beta_3)^2$\\$(\mu_{\bar{d}}-\mu_{\bar{u}})^2/9$\end{tabular}} &&92.91&\\
		$T_{bc}^{(1,0)}(7528)\rightarrow T_{bc}^{(0,1)}(7320)\gamma$&&& 52.96&\\
		$T_{bc}^{(1,0)}(7528)\rightarrow T_{bc}^{(0,1)}(7208)\gamma$&&& 3.28&\\
		$T_{bc}^{(1,0)}(7342)\rightarrow T_{bc}^{(0,1)}(7320)\gamma$&&& 0.02&\\
		$T_{bc}^{(1,0)}(7342)\rightarrow T_{bc}^{(0,1)}(7208)\gamma$&&& 54.37&\\\cline{2-5}
		$T_{bc}^{(0,0)}(7322)\rightarrow T_{bc}^{(0,1)}(7320)\gamma$&\multirow{2}{*}{\begin{tabular}{c}$\frac19[(\sqrt{3}(\alpha_1\beta_3-3\alpha_2\beta_2))(\mu_b-\mu_c)+$\\$\sqrt{6}\alpha_1\beta_1(\mu_{\bar{d}}+\mu_{\bar{u}}-\mu_b-\mu_c)]^2$\end{tabular}} && {$2.25\times 10^{-5}$}&\\
		$T_{bc}^{(0,0)}(7322)\rightarrow T_{bc}^{(0,1)}(7208)\gamma$&&& {3.04}&\\
		\hline\hline
	\end{tabular}	
\end{table}

\subsubsection{The $bc \bar{q}\bar{q}^\prime$ systems}

\begin{table}[!h]
	\caption{Radiative transitions of $bc\bar{s}\bar{s}$ states in units of keV. A tetraquark is denoted as $T^{(I,J)}(Mass)$. In the second column, $\alpha_i$ and $\beta_i$ represent the weight coefficients of the spin-color configurations in the initial and final states, respectively.}\label{tab:bcss-radiative-width}
	\begin{tabular}{c|c|cccccc}\hline\hline
		\multirow{2}{*}{Radiative decay channel}&\multicolumn{2}{c}{$\Gamma$ (keV)}\\\cline{2-3}
		&$\langle\mu\rangle^2$&$I_3=0$\\\hline
		$T_{bc}^{(0,2)}(7618)\rightarrow T_{bc}^{(0,1)}(7586)\gamma$&\multirow{2}{*}{\begin{tabular}{c}$\frac56[\beta_1(\mu_b+\mu_c-2\mu_{\bar{s}})+$\\$\sqrt{2}\beta_3(\mu_b-\mu_c)]^2$\end{tabular}}&0.07\\
		$T_{bc}^{(0,2)}(7618)\rightarrow T_{bc}^{(0,1)}(7510)\gamma$& &0.05\\\cline{2-3}
		$T_{bc}^{(0,1)}(7641)\rightarrow T_{bc}^{(0,2)}(7618)\gamma$&$\begin{tabular}{c}$\frac56[\alpha_1(\mu_b+\mu_c-2\mu_{\bar{s}})+$\\$\sqrt{2}\alpha_3(\mu_b-\mu_c)]^2$\end{tabular}$&$1.02\times10^{-5}$\\\cline{2-3}
		$T_{bc}^{(0,1)}(7641)\rightarrow T_{bc}^{(0,1)}(7586)\gamma$&\multirow{3}{*}{\begin{tabular}{c}$\frac12[\alpha_1\beta_1(\mu_b+\mu_c+2\mu_{\bar{s}})+$\\$2\alpha_2\beta_2(\mu_b+\mu_c)+3\alpha_3\beta_3\mu_{\bar{s}}+$\\$\sqrt{2}(\alpha_3\beta_1+\alpha_1\beta_3)(\mu_c-\mu_b)]^2$\end{tabular}}&0.07\\
		$T_{bc}^{(0,1)}(7641)\rightarrow T_{bc}^{(0,1)}(7510)\gamma$& &1.05\\
		$T_{bc}^{(0,1)}(7586)\rightarrow T_{bc}^{(0,1)}(7510)\gamma$& &0.21\\\cline{2-3}
		$T_{bc}^{(0,1)}(7641)\rightarrow T_{bc}^{(0,0)}(7483)\gamma$&\begin{tabular}{c}$\frac19[\sqrt{6}\alpha_1\beta_1(\mu_b+\mu_c-2\mu_{\bar{s}})-$\\$(\sqrt{3}\alpha_3\beta_1-3\alpha_2\beta_2)(\mu_b-\mu_c)]^2$\end{tabular}&0.12\\\cline{2-3}
		$T_{bc}^{(0,0)}(7669)\rightarrow T_{bc}^{(0,1)}(7641)\gamma$&\multirow{3}{*}{\begin{tabular}{c}$\frac19[\sqrt{6}\alpha_1\beta_1(\mu_b+\mu_c-2\mu_{\bar{s}})-$\\$(\sqrt{3}\alpha_1\beta_3-3\alpha_2\beta_2)(\mu_b-\mu_c)]^2$\end{tabular}}&0.05\\
		$T_{bc}^{(0,0)}(7669)\rightarrow T_{bc}^{(0,1)}(7586)\gamma$& &0.62\\
		$T_{bc}^{(0,0)}(7669)\rightarrow T_{bc}^{(0,1)}(7510)\gamma$& &0.69\\
		\hline\hline
	\end{tabular}	
\end{table}

We present the radiative decays for the $bc\bar{n}\bar{n}^\prime$   and $bc\bar{s}\bar{s}$ systems in Tables \ref{tab:bcnn-radiative-width} and \ref{tab:bcss-radiative-width}, respectively.
The isospin of the compact $bc\bar{n}\bar{n}^\prime$ states can be 1 or 0. Their radiative transitions occur not only between states with different total spins but also between different isospin multiplet states. This offers a useful handle for experimental detection through radiative decay links.

The $bc\bar{n}\bar{n}^\prime$ system has fifty-one radiative decay channels. Numerous radiative decay channels depend solely on the spin-flip transition of the initial and final anti-diquarks.
The electromagnetic widths span several orders of magnitude, ranging from $10^{-5}$ to several hundred keV.
Notably, two channels exhibit widths exceeding $100$ keV: $T_{bc}^{(1,2)}(7468)\rightarrow T_{bc}^{(0,1)}(7208)\gamma$ ($244.15$ keV) and $T_{bc}^{(1,1)}(7438)\rightarrow T_{bc}^{(0,0)}(7151)\gamma$ ($114.53$ keV).
These channels are likely to be important in the experimental search for $bc\bar{n}\bar{n}^\prime$ tetraquark states.

In the $bc\bar{s}\bar{s}$ system, the amplitude expressions in Table \ref{tab:bcss-radiative-width} show that the configurations of the initial- and final-state wave functions directly affect the radiative decay widths through the weight coefficients $\alpha_i$ and $\beta_i$.
The calculated widths of the ten radiative decay channels exhibit an enormous variation, spanning five orders of magnitude from $10^{-5}$ keV to over 1 keV, which makes a challenge for experimental measurements.

\section{SUMMARY}\label{sec4}

In this work, we systematically study magnetic moments of $S$-wave $QQ^\prime  \bar{q}\bar{q}^\prime$ tetraquarks using wave functions obtained in a mass splitting model.
We also define magnetic coupling matrices for the considered $J^P=1^+$ systems in both compact and molecule pictures, with which the minimum and maximum magnetic moments of a system can be calculated. This notion actually leads to an alternative method for the systematic calculation of the multiquark MM.

Our study indicates that it is impossible to distinguish whether the $J^P=2^+$ $QQ^\prime  \bar{q}\bar{q}^\prime$ states are compact tetraquarks or molecular states from the magnetic moment alone.
The magnetic moments of the $J^P=1^+$ $bc\bar{q}\bar{q}^\prime$ and $J^P=1^+$ $QQ\bar{n}\bar{s}$ states are influenced by the diquark-spin mixing within the compact tetraquark picture. Those magnetic moments can reflect the structural differences between molecule and compact tetraquark configurations. 
For the $J^P=0^+$ states, their magnetic moments are all zero because of the spin $S=0$. 
Other physical quantities to probe inner structures of the $2^+$ and $0^+$ states are needed. The radiative transitions between different tetraquarks are one such quantity.

For the $cc\bar{n}\bar{n}^\prime$ and $bb\bar{n}\bar{n}^\prime$ systems, the magnetic moments of the $I(J^P)=1(1^+)$ states in both molecular and compact tetraquark pictures are identical, whereas those of the $I(J^P)=0(1^+)$ states are determined by the weights of the spin-color wave functions, offering a probe of their internal configurations.
Besides the spectrum and strong decays, the radiative decays of the $I(J^P)=1(2^+)$, $1(1^+)$, and $1(0^+)$ states into another tetraquarks can also be used to gain insight into their substructures.
The obtained radiative decay widths range from $6.70\times10^{-6}$ to $306.37$ keV ($cc\bar{n}\bar{n}^\prime$) and $0.01$ to $293.99$ keV ($bb\bar{n}\bar{n}^\prime$). 
The $I_3=-1$ to $I_3=+1$ width ratio for states in the same iso-triplet is about $13.60$ ($cc\bar{n}\bar{n}^\prime$) and $3.00$ ($bb\bar{n}\bar{n}^\prime$). 
If the $T_{cc}(3875)^+$ corresponds to the low-mass $I(J^P)=0(1^+)$ compact tetraquark, its magnetic moment is predicted to be $0.45$ $\mu_N$, and five higher tetraquarks can radiatively decay into this exotic state, with one prominent channel: $T_{cc}^{(1,2)}(4143)\rightarrow T_{cc}^{(0,1)}(3878)\gamma$ ($306.37$ keV).

For the $cc\bar{s}\bar{s}$ and $bb\bar{s}\bar{s}$ states, the magnetic moment (a few keV) cannot be used to identify their structures, molecules or compact tetraquarks, but the radiative transitions between $J=1$ and $J=0$ can tell us the proportion of the $S_{QQ}=1$ state in the scalar tetraquarks. 

For the $cc\bar{n}\bar{s}$, $bb\bar{n}\bar{s}$, and $bc\bar{n}\bar{s}$ systems, the magnetic moments and radiative decays involve couplings between different light anti-diquark spin configurations.  
The magnetic moments of $bb\bar{n}\bar{s}$ states are generally smaller than their $cc\bar{n}\bar{s}$ analogs. 
However, the magnetic monent of $T_{bb}^{(\frac12,1)}(10819)$ ($1.53$ $\mu_N$) exceeds that of $T_{cc}^{(\frac12,1)}(4147)$ ($1.20$ $\mu_N$), indicating the complex dynamics in tetraquark states. For the  $QQ\bar{n}\bar{s}$ states, we observe that the radiative decay widths of $I_3=-\frac12$ case are generally larger than those of $I_3=+\frac12$ case. However, this trend is broken in some deacy channels of the $bb\bar{n}\bar{s}$ and $bc\bar{n}\bar{s}$ systems where $I_3=1/2$ dominates (see Tables \ref{tab:QQns-radiative-width} and \ref{tab:ratio-bcns}).

For the $bc\bar{n}\bar{n}^\prime$ system, although the isospin triplet and singlet states share identical magnetic moment formulas, their magnetic moments differ due to the distinct weights $\alpha_i$ in the tetraquark wave functions.

To summarize, the obtained magnetic moments and radiative decays of the $T_{cc}(3875)^+$ and its partner states provide a perspective on the nature of doubly heavy tetraquark states, as these observables are sensitive to the internal structures. The study also offers a method to explore the relationships among exotic states through radiative decays, which could aid experimental searches for new doubly heavy tetraquark states.

\section*{Acknowledgments}
We are grateful for the helpful discussions with Dr. Hao-Song Li. This project was supported by the National Natural Science Foundation of China under Grant Nos. 12475143, 12235008, 12405103, 12405160, and 11905114 and by
the Shandong Province Natural Science Foundation (ZR2023MA041).

\end{document}